\documentclass{aa}
\usepackage{graphicx}
\begin{document}

\title{The angular sizes of dwarf stars and subgiants}
\subtitle{Surface brightness relations calibrated by interferometry}
\author{P. Kervella \inst{1,2},
        F. Th\'evenin\inst{3},
        E. Di Folco\inst{4}
        \and
       D. S\'egransan\inst{5}}
\offprints{P. Kervella}
\institute{
LESIA, UMR 8109, Observatoire de Paris-Meudon, 5, place Jules Janssen, F-92195 Meudon Cedex, France
\and European Southern Observatory,
Alonso de Cordova 3107, Casilla 19001, Vitacura, Santiago 19, Chile
\and Observatoire de la C\^ote 
   d'Azur, BP 4229, F-06304 Nice Cedex 4, France
\and European Southern Observatory, Karl-Schwarzschild-str. 2,
D-85748 Garching, Germany
\and Observatoire de Gen\`eve, CH-1290 Sauverny, Switzerland
}
\titlerunning{The angular sizes of dwarf stars and subgiants}
\authorrunning{P. Kervella et al.}
\mail{Pierre.Kervella@obspm.fr}
\date{Received date / Accepted date}

\abstract{
The availability of a number of new interferometric measurements of
Main Sequence and subgiant stars makes it possible to
calibrate the surface brightness relations of these stars
using exclusively direct angular diameter measurements. These empirical laws make it possible to
predict the limb darkened angular diameters $\theta_{\rm LD}$ of dwarfs and subgiants
using their dereddened Johnson magnitudes, or their effective temperature.
The smallest intrinsic dispersions of $\sigma \le 1\,\%$ in $\theta_{\rm LD}$ are obtained
for the relations based on the $K$ and $L$ magnitudes, for instance
$\log \theta_{\rm LD} = 0.0502\,(B-L) + 0.5133 - 0.2\,L$ or $\log \theta_{\rm LD} = 0.0755\,(V-K) + 0.5170 - 0.2\,K$.
Our calibrations are valid between the spectral types A0 and M2 for dwarf stars
(with a possible extension to later types when using the effective temperature),
and between A0 and K0 for subgiants.
Such relations are particularly useful for estimating the angular sizes of calibrators for
long-baseline interferometry from readily available broadband photometry.
}

\maketitle

\section{Introduction}\label{sec:int}
The surface brightness (hereafter SB) relations link the emerging flux per
solid angle unit of a light-emitting body to its color, or effective temperature.
These relations are of considerable astrophysical interest, as a well-defined
relation between a particular color index and the surface brightness can provide
accurate predictions of the stellar angular diameters. Such predictions are
essential for the calibration of long-baseline interferometric observations.
We propose in the present paper new and accurate calibrations of the SB-color
relations of dwarfs and subgiants based on direct interferometric measurements
of nearby members of these two luminosity classes.
Our primary purpose is to establish reliable relations that can be used to predict
the angular sizes of calibrator stars for long-baseline interferometry.

After defining the surface brightness relations (Sect.~\ref{def_relations}),
we discuss in Sect.~\ref{section_sample} the sample of measurements that
we selected for our calibrations (interferometric and photometric data).
Sect.~\ref{section_relations} is dedicated to the calibration of the empirical SB
relations, relative to the color indices and to the effective temperature,
for stars of spectral types A0 to M2. We also derive inverse relations to estimate the
effective temperature from broadband photometry and angular diameter measurements.
As the established relations are intended to be used primarily to predict angular diameters,
we discuss in Sect.~\ref{estim_errors} their associated errors in this context.
In Sect.~\ref{comparison_section}, we search for a possible
instrumental bias linked to one of the five interferometric instruments
represented in our sample.
Numerous versions of the SB relations have been
established in the literature, mostly for giants and supergiants, and we discuss them
in Sect.~\ref{compare_rel}.
Main Sequence stars are potentially very good calibrators for long-baseline interferometry,
and we discuss this particular application of our SB relations in Sect.~\ref{calib_interf}.

\section{Direct and inverse surface brightness relations \label{def_relations}}

By definition, the bolometric surface flux
$f \sim L / D^2$ is linearly proportional to $T_{\rm eff}^4$, where $L$ is the bolometric
flux of the star, $D$ its bolometric diameter and $T_{\rm eff}$ its effective temperature.
In consequence, $F = \log f$ is a linear function of the stellar color indices expressed
in magnitudes (logarithmic scale), and SB relations can be fitted using
(for example) the following expressions:
\begin{equation}\label{be_eq}
F_{B} = a_0\,(B-V)_0 + b_0
\end{equation}
\begin{equation}
F_{V} = a_1\,(V-K)_0 + b_1
\end{equation}
\begin{equation}
F_{H} = a_2\,(B-H)_0 + b_2
\end{equation}
where $F_{\lambda}$ is the surface brightness.
When considering a perfect blackbody curve, any color can in principle
be used to obtain the SB.
The index $0$ designates the dereddened magnitudes,
and the $a_i$ and $b_i$ coefficients represent respectively the slopes and zero points
of the different versions of the SB relation. The particular expression of the SB relation
$F_{V} (V-R)$ is also known as the Barnes-Evans (B-E) relation, and is historically the first version
to have been calibrated empirically (Barnes, Evans \& Pearson~\cite{barnes76}).
However, the relatively large intrinsic dispersion of the visible B-E relation
has led many authors to prefer its infrared counterparts, in particular those based
on the $K$ band magnitudes ($\lambda = 2.0-2.4\ \mu$m), as
infrared wavelengths are less affected by interstellar extinction.
The surface brightness $F_\lambda$ is given by the following expression
(Fouqu\'e \& Gieren ~\cite{fouque97}) :
\begin{equation}\label{f_eq}
F_{\lambda} = 4.2207 - 0.1\,m_{\lambda_0} - 0.5\,\log \theta_{\rm LD}
\end{equation}
where $ \theta_{\rm LD}$ is the limb darkened angular diameter,
i.e. the angular size of the stellar photosphere.

The linear expressions of the SB can be inverted easily to
predict angular diameters, and give linear relations such as:
\begin{equation}
\log \theta_{\rm LD} = c_1\,(V-K) + d_1 - 0.2\,V
\end{equation}
for the $F_V(V-K)$ inversion. We have in this example:
\begin{equation}
c_1 = -2\,a_1
\end{equation}
\begin{equation}
d_1 = 2\,(4.2207 - b_1)
\end{equation}
In the present paper, we will refer to both the direct and inverse relations as
``SB relations''.

\section{Selected measurement sample \label{section_sample}}

\subsection{Angular diameters}

Over the past two years, sixteen new angular diameter measurements of nearby
Main Sequence and subgiant stars were obtained with the VLT Interferometer
(Glindemann et al.~\cite{glindemann00}; \cite{glindemann03a}; \cite{glindemann03b})
equipped with the fiber-based beam
combiner VINCI (Kervella et al.~\cite{kervella00}; \cite{kervella03a}).
To complement this sample, we have searched the literature, and added to our list the
measurements related to the stars of luminosity classes IV and V.
Most of the visible and infrared interferometers are represented in our sample,
with measurements from the
NII ({\it Narrabri Intensity Interferometer}; Hanbury Brown, Davis \& Allen~\cite{hanbury67}),
the Mk\,III (Shao et al.~\cite{shao88}),
the PTI ({\it Palomar Testbed Interferometer}; Colavita et al.~\cite{colavita99})
and the NPOI ({\it Navy Prototype Optical Interferometer}; Armstrong et al.~\cite{armstrong98}).
Our findings originally included a few lunar occultation measurements, but
they were rejected as they were related to variable or multiple stars,
or their precision was not sufficient to give them any weight in the fitting process.

To obtain a consistent sample of limb darkened (LD) angular diameters
we have retained solely the uniform disk (UD) values from the literature.
The conversion of these model-independent measurements to LD values
was achieved using the linear LD coefficients $u$ from Claret~(\cite{claret00}),
and the conversion formula from Hanbury Brown et al.~(\cite{hanbury74a}).
These coefficients are broadband, single-parameter approximations of the
Kurucz~(\cite{kurucz92}) model atmospheres. They are tabulated for a grid of temperatures,
metallicities and surface gravities and we have chosen the closest models to the
physical properties of the stars. We have considered a uniform microturbulent velocity
of 2\,km.s$^{-1}$ for all stars. This single source for limb darkening corrections
ensures the self-consistency of our final sample.

\subsection{Photometry}

All the apparent magnitudes that we have retained from the literature are expressed in the
Johnson system.
When available, we have preferentially kept the uncertainties given by the original authors,
otherwise we adopted arbitrarily a conservative error bar.
The $U$ band magnitudes were obtained from Morel et al.~(\cite{morel78}) and
Mermilliod~(\cite{mermilliod86}), and we adopted a $\pm 0.02$ error.
The $B$, $V$, $R$ and $I$ bands were obtained from several online catalogues available
through VIZIER (Ochsenbein, Bauer \& Marcout~\cite{ochsenbein00}),
and we also adopted a $\pm 0.02$ uncertainty.
For the $J$ to $L$ infrared bands, references are not so easy to find, as many bright stars
are unfortunately absent from the recent infrared surveys, like 2MASS (Cutri et al.~\cite{cutri03})
or DENIS (Fouqu\'e et al.~\cite{fouque00}).
We have relied on the VIZIER database to obtain the infrared magnitudes of
our sample of stars. In some cases, the references we used
are 30 years old, but many of them have small and reliable uncertainties.
The original references of the measurements are given in the footnotes of Table~\ref{photom_table}.

\subsection{Data selection \label{selection}}

The SB relations rely on the assumption that stars behave
like black bodies, i.e. that their colors are mainly governed by their
effective temperature.
A severe deviation from this assumption will cause a discrepancy
between the actual flux per surface unit and the temperature of the star.

For instance, if there is a second, unresolved star near the main
object, its additional flux will bias the spectral
energy distribution. For this reason, we have rejected the binary and multiple
objects for which separate photometry of the components is not available.

The presence of warm material in the circumstellar environment can also create
an excess at infrared wavelengths. While this signature is most useful for
identifying the stars surrounded by protoplanetary disks, it creates a
bias in the measured color of the star. Some of the stars we selected are surrounded
by debris disks ($\epsilon$\,Eri, $\alpha$\,PsA, $\tau$\,Cet, $\beta$\,Leo), but the
contribution of the circumstellar material is negligible, even in the infrared
$J$ to $L$ bands. The material surrounding these stars is very cold and radiates
mostly in the far infrared domain. We have rejected the measurement
of $\beta$\,Pic from Di Folco et al.~(\cite{difolco04}), due to its large uncertainty
($\simeq 10\%$) and to the relatively high density of its edge-on circumstellar
disk that could cause significant extinction.

The fast rotating stars can deviate significantly from the black body assumption.
As demonstrated by the VINCI observations of the nearby Be
star $\alpha$\,Eri (Domiciano de Souza et al.~\cite{domiciano03}), the photosphere
of these objects can be deformed by their fast rotation.
This creates differential limb darkening between the pole and the equator
which appear to have different effective temperatures. This makes it particularly
difficult to define the true photometric solid angle subtended by these objects.
In addition, many fast rotating stars go through episodes of mass loss, that are likely
to create a warm circumstellar environment. The presence of such hot material around
the star will create a bias in the flux and color of the star. For these
reasons, we have chosen to reject the known fast rotators ($v \sin i \ge 100$\,km.s$^{-1}$)
and the Be stars from our list.

The very low mass stars {\it Proxima} (GJ\,551, M5.5V) and {\it Barnard's star} (GJ\,699, M4V)
have been excluded from our fitting procedures for three reasons.
The first is that because of their very low effective
temperatures the molecular absorption bands dominate their spectra and lead to a
significant discrepancy with the hotter stars. Second, these stars are variable,
presenting occasional flares that make it difficult to estimate their magnitudes.
Third, they present chromospheric activity that could bias their magnitudes in the
$U$ to $V$ colors. However, we have kept these two stars on the SB relation
figures for comparison.

The spectroscopic and eclipsing binaries
are less useful for the estimation of the surface brightness relation,
as it is in general impossible to measure separately the magnitudes
of these stars with the required precision.
For this reason, we have not included in our sample
the angular diameter measurements obtained by spectroscopic
or photometric methods.
For the interested reader, a rather complete compilation of the
measurements using these techniques can be found in the CADARS catalogue
by Pasinetti-Fracassini et al.~(\cite{pasinetti01}).

\subsection{Final sample \label{final_sample}}

We report in Tables~\ref{angdiams_table_V} and \ref{angdiams_table_IV} the complete set of
measurements that we have considered for our fit. In this table,
the angular diameters $\theta_{\rm UD}$ (uniform disk)
and $\theta_{\rm LD}$ (limb-darkened disk) are expressed in milliarcseconds (mas).
The limb darkening conversion coefficient $k = \theta_{\rm LD}/\theta_{\rm UD}$ was computed for each star
based on the tables of Claret~(\cite{claret00}). When a physical parameter was not available in the literature,
it has been estimated approximately, and appears in {\it italic} characters. The observation wavelength
$\lambda$ is given as either the name of the photometric band ($V$, $H$, $K$) or the actual
wavelength in $\mu$m.
The error bar in the angular diameter of the Sun (G2V) has been set arbitrarily to $\pm 0.1$\%.
The parallaxes are from the {\it Hipparcos} catalogue (Perryman et al.~\cite{hip}), except the $\alpha$\,Cen
value that was taken from S\"oderhjelm~(\cite{soderhjelm99}), who derived it from reprocessed
{\it Hipparcos} data. The interferometer used for each measurement is indicated in the "Instr." column.

\begin{table*}
\caption{Angular diameters of dwarf stars (luminosity class V) measured
by long-baseline interferometry (apart from the Sun). They are expressed
in mas, and $T_{\rm eff}$ is in K. "Ref.$_1$" designates the reference used for $T_{\rm eff}$, $\log g$
and [Fe/H]. When unavailable, the metallicity has been set arbitrarily to the solar value.
"Ref.$_2$" designates the reference used for each angular diameter measurement
(expressed in mas). The errors are given in superscript close to each value.}
\label{angdiams_table_V}
\begin{tabular}{llrcccclccccc}
\hline
Star & Spect. & $\pi$\,(mas) & Ref.$_1$& $T_{\rm eff}$ & $\log g$ & [Fe/H]&
Instr. & Ref.$_2$ & $\lambda$ & $\theta_{\rm UD}$ & $k$ & $\theta_{\rm LD}$ \\
\hline \noalign{\smallskip}
$\alpha$\,Lyr&A0V& $128.93^{0.55}$ &(a,e) &9522&3.98&-0.33&PTI&(8)& $K$ & $3.24^{0.01}$ &1.012& $3.28^{0.01}$ \\
$\alpha$\,Lyr&A0V& $128.93^{0.55}$ &(a,e) &9522&3.98&-0.33&NII&(1) & $V$ & $3.08^{0.07}$ &1.046& $3.22^{0.07}$ \\
$\alpha$\,Lyr&A0V& $128.93^{0.55}$ &(a,e) &9522&3.98&-0.33&Mk\,III&(4) & $0.8$ & $3.15^{0.03}$ &1.028& $3.24^{0.03}$ \\
$\alpha$\,Lyr&A0V& $128.93^{0.55}$ &(a,e) &9522&3.98&-0.33&Mk\,III&(4) & $0.55$ & $3.00^{0.05}$ &1.046& $3.13^{0.05}$ \\
$\alpha$\,CMa\,A&A1V& $379.21^{1.58}$ &(b) &9800&4.30&0.40&NII&(16) & $V$ & $5.60^{0.07}$ &1.045& $5.85^{0.07}$ \\
$\alpha$\,CMa\,A&A1V& $379.21^{1.58}$ &(b) &9800&4.30&0.40&VLTI&(9) & $K$ & $5.94^{0.02}$ &1.012& $6.01^{0.02}$ \\
$\alpha$\,CMa\,A&A1V& $379.21^{1.58}$ &(b) &9800&4.30&0.40&Mk\,III&(4) & $0.8$ & $5.82^{0.11}$ &1.027& $5.98^{0.11}$ \\
$\beta$\,Leo&A3V& $90.16^{0.89}$ &(g) &8570&4.26& 0.20 &VLTI&(10) & $K$ & $1.43^{0.03}$ &1.015& $1.45^{0.03}$ \\
$\beta$\,Leo&A3V& $90.16^{0.89}$ &(g) &8570&4.26& 0.20 &NII&(1) & $V$ & $1.25^{0.09}$ &1.052& $1.31^{0.09}$ \\
$\alpha$\,PsA&A3V& $130.08^{0.92}$ &(g) &8760&4.22&0.43&NII&(1) & $V$ & $1.98^{0.13}$ &1.050& $2.08^{0.14}$ \\
$\alpha$\,PsA&A3V& $130.08^{0.92}$ &(g) &8760&4.22&0.43&VLTI&(10) & $K$ & $2.20^{0.02}$ &1.014& $2.23^{0.02}$ \\
$\alpha$\,Cen\,A&G2V& $747.10^{1.20}$ &(k) &5790&4.32&0.20&VLTI&(13) & $K$ & $8.31^{0.02}$ &1.024& $8.51^{0.02}$ \\
{\it Sun}&G2V& $^{}$ &&5770&&&&&&&& $1919260^{0.1\%}$ \\
$\tau$\,Cet&G8V& $274.18^{0.80}$ &(i) &5400&4.55&-0.40&VLTI&(10) & $K$ & $2.03^{0.03}$ &1.024& $2.08^{0.03}$ \\
GJ\,166\,A&K1V& $198.25^{0.84}$ &(a) &5073&4.19&-0.31&VLTI&(15) & $K$ & $1.60^{0.06}$ &1.029& $1.65^{0.06}$ \\
$\alpha$\,Cen\,B&K1V& $747.10^{1.20}$ &(k) &5260&4.51&0.23&VLTI&(13) & $K$ & $5.86^{0.03}$ &1.026& $6.01^{0.03}$ \\
$\epsilon$\,Eri&K2V& $310.74^{0.85}$ &(a) &5052&4.57&-0.15&VLTI&(10) & $K$ & $2.09^{0.03}$ &1.027& $2.15^{0.03}$ \\
GJ\,105\,A&K3V& $138.72^{1.04}$ &(a) &4718&4.50&-0.07&PTI&(12) & $H, K$ & $0.91^{0.07}$ &1.032& $0.94^{0.07}$ \\
GJ\,570\,A&K4V& $169.31^{1.67}$ &(a) &4533&4.79&0.02&VLTI&(15) & $K$ & $1.19^{0.03}$ &1.030& $1.23^{0.03}$ \\
$\epsilon$\,Ind\,A&K4.5V& $275.49^{0.69}$ &(b) &4500&4.50&-0.10&VLTI&(15) & $K$ & $1.84^{0.02}$ &1.030& $1.89^{0.02}$ \\
GJ\,380&K7V& $205.23^{0.81}$ &(a) &3861&4.68&-0.93&PTI&(12) & $H, K$ & $1.27^{0.04}$ &1.018& $1.29^{0.04}$ \\
GJ\,191&M1V& $255.12^{0.86}$ &(b) &3524&4.87&-0.50&VLTI&(14) & $K$ & $0.68^{0.06}$ &1.016& $0.69^{0.06}$ \\
GJ\,887&M0.5V& $303.89^{0.87}$ &(f) &3645&4.80& {\it0.00} &VLTI&(14) & $K$ & $1.37^{0.04}$ &1.018& $1.39^{0.04}$ \\
GJ\,205&M1.5V& $175.72^{1.20}$ &(b) &3626&4.80&0.60&VLTI&(14) & $K$ & $1.12^{0.11}$ &1.020& $1.15^{0.11}$ \\
GJ\,15\,A&M2V& $280.26^{1.05}$ &(a) &3721&5.00&-1.40&PTI&(12) & $H, K$ & $0.98^{0.05}$ &1.017& $1.00^{0.05}$ \\
GJ\,411&M1.5V& $392.52^{0.91}$ &(h) &3620&4.90&-0.20&PTI&(12) & $H, K$ & $1.41^{0.03}$ &1.019& $1.44^{0.03}$ \\
GJ\,699&M4Ve& $549.30^{1.58}$ &(a) &3201&5.00&-0.90&PTI&(12) & $H, K$ & $0.99^{0.04}$ &1.018& $1.00^{0.04}$ \\
{\it Proxima}&M5.5V& $772.33^{2.42}$ &(f) &3006&5.19&{\it 0.00}&VLTI&(14) & $K$ & $1.02^{0.08}$ &1.030& $1.05^{0.08}$ \\
\hline
\end{tabular}
\begin{itemize}{}{}
\item Ref.$_1$ for $T_{\rm eff}$, $\log g$ and [Fe/H]:
(a) Cenarro et al.~(\cite{cenarro01});
(b) Cayrel de Strobel et al.~(\cite{cayrel97});
(c) Allende Prieto et al.~(\cite{allende02};
(d) Gray et al.~(\cite{gray01});
(e) Th\'evenin \& Idiart~(\cite{thevenin99});
(f) S\'egransan et al.~(\cite{segransan03});
(g) Erspamer \& North~(\cite{erspamer03});
(h) Cayrel de Strobel et al.~(\cite{cayrel01});
(i) Di Folco et al.~(\cite{difolco04});
(j) Morel et al.~(\cite{morel01});
(k) Morel et al.~(\cite{morel00}).
\item Ref.$_2$ for angular diameters:
(1) Hanbury Brown, Davis \& Allen~(\cite{hanbury74b});
(2) Kervella et al.~(\cite{kervella04a});
(3) Nordgren et al.~(\cite{nordgren01});
(4) Mozurkewich et al.~(\cite{mozurkewich03});
(5) Th\'evenin et al.~(\cite{thevenin04});
(6) Boden et al.~(\cite{boden98});
(7) Nordgren et al.~(\cite{nordgren99});
(8) Ciardi et al.~(\cite{ciardi01});
(9) Kervella et al.~(\cite{kervella03b});
(10) Di Folco et al.~(\cite{difolco04});
(11) Nordgren et al.~(\cite{nordgren01});
(12) Lane et al.~(\cite{lane01});
(13) Kervella et al.~(\cite{kervella03c});
(14) S\'egransan et al.~(\cite{segransan03});
(15) S\'egransan et al.~(\cite{segransan04});
(16) Davis et al.~(\cite{davis86}).
\end{itemize}
\end{table*}

\begin{table*}
\caption{Angular diameters of subgiant stars (luminosity class IV) measured by interferometry.
The references and notations are given in Table~\ref{angdiams_table_V}.}
\label{angdiams_table_IV}
\begin{tabular}{llrcccclccccc}
\hline
Star & Spect. & $\pi$\,(mas) & Ref.$_1$& $T_{\rm eff}$ & $\log g$ & [Fe/H]&
Instr. & Ref.$_2$ & $\lambda$ & $\theta_{\rm UD}$ & $k$ & $\theta_{\rm LD}$ \\
\hline \noalign{\smallskip}
$\gamma$\,Gem&A0IV& $31.12^{2.33}$ &(b) &9260&3.60&-0.12&NII&(1) & $V$ & $1.32^{0.09}$ &1.047& $1.38^{0.09}$ \\
$\alpha$\,CMi\,A&F5IV-V& $285.93^{0.88}$ &(c) &6530&3.96&-0.05&VLTI&(2) & $K$ & $5.38^{0.05}$ &1.019& $5.48^{0.05}$ \\
$\alpha$\,CMi\,A&F5IV-V& $285.93^{0.88}$ &(c) &6530&3.96&-0.05&NPOI&(11) & $V$ & $5.19^{0.04}$ &1.057& $5.49^{0.04}$ \\
$\alpha$\,CMi\,A&F5IV-V& $285.93^{0.88}$ &(c) &6530&3.96&-0.05&Mk\,III&(4) & $0.8$ & $5.32^{0.08}$ &1.039& $5.53^{0.08}$ \\
$\alpha$\,CMi\,A&F5IV-V& $285.93^{0.88}$ &(c) &6530&3.96&-0.05&Mk\,III&(4) & $0.55$ & $5.30^{0.11}$ &1.057& $5.60^{0.11}$ \\
$\eta$\,Boo&G0IV& $88.17^{0.75}$ &(a) &6003&3.62&0.25&VLTI&(5) & $K$ & $2.15^{0.03}$ &1.022& $2.20^{0.03}$ \\
$\eta$\,Boo&G0IV& $88.17^{0.75}$ &(a) &6003&3.62&0.25&Mk\,III&(4) & $0.8$ & $2.18^{0.02}$ &1.044& $2.27^{0.03}$ \\
$\eta$\,Boo&G0IV& $88.17^{0.75}$ &(a) &6003&3.62&0.25&Mk\,III&(4) & $0.55$ & $2.13^{0.03}$ &1.063& $2.26^{0.03}$ \\
$\eta$\,Boo&G0IV& $88.17^{0.75}$ &(a) &6003&3.62&0.25&NPOI&(11) & $V$ & $2.17^{0.06}$ &1.064& $2.31^{0.06}$ \\
$\zeta$\,Her\,A&G0IV& $92.64^{0.60}$ &(j) &5820&3.85&0.04&Mk\,III&(4) & $0.8$ & $2.26^{0.05}$ &1.045& $2.36^{0.05}$ \\
$\zeta$\,Her\,A&G0IV& $92.64^{0.60}$ &(j) &5820&3.85&0.04&Mk\,III&(4) & $0.55$ & $2.13^{0.03}$ &1.065& $2.27^{0.03}$ \\
$\zeta$\,Her\,A&G0IV& $92.64^{0.60}$ &(j) &5820&3.85&0.04&NPOI&(11) & $V$ & $2.37^{0.08}$ &1.065& $2.52^{0.09}$ \\
$\mu$\,Her&G5IV& $119.05^{0.62}$ &(a) &5411&3.87&0.16&Mk\,III&(4) & $0.8$ & $1.86^{0.04}$ &1.049& $1.95^{0.04}$ \\
$\mu$\,Her&G5IV& $119.05^{0.62}$ &(a) &5411&3.87&0.16&Mk\,III&(4) & $0.55$ & $1.81^{0.03}$ &1.070& $1.93^{0.03}$ \\
$\beta$\,Aql&G8IV& $72.95^{0.83}$ &(a) &5041&3.04&-0.04&NPOI&(7)& $V$ & $2.07^{0.09}$ &1.075& $2.23^{0.10}$ \\
$\eta$\,Cep&K0IV& $69.73^{0.49}$ &(a) &5013&3.19&-0.19&NPOI&(7)& $V$ & $2.51^{0.04}$ &1.064& $2.67^{0.04}$ \\
$\delta$\,Eri&K0IV& $110.58^{0.88}$ &(a) &4884&3.40&-0.11&VLTI&(5) & $K$ & $2.33^{0.03}$ &1.027& $2.39^{0.03}$ \\
\hline
\end{tabular}
\end{table*}

\begin{table*}
\caption{
Apparent magnitudes of the dwarf stars (upper part) and subgiants (lower part) of our sample.
The uncertainty adopted for each apparent magnitude value is given in superscript.
}
\label{photom_table}
\begin{tabular}{lrrrrrrrrr}
\hline
Star & $m_U$$^{(a)}$ & $m_B$$^{(b)}$ & $m_V$$^{(b)}$ & $m_R$$^{(c)}$ & $m_I$$^{(c)}$ & $m_J$$^{(d)}$ & $m_H$$^{(d)}$ & $m_K$$^{(d)}$ & $m_L$$^{(d)}$ \\
\hline
\noalign{\smallskip}
\object{$\alpha$\,Lyr} & $0.03^{0.02}$ & $0.03^{0.02}$ & $0.03^{0.02}$ & $0.04^{0.02}$ & $0.03^{0.02}$ & $0.00^{0.02}$ & $0.00^{0.01}$ & $0.00^{0.01}$ & $0.00^{0.01}$\\
\object{$\alpha$\,CMa\,A} & -$1.51^{0.02}$ & -$1.46^{0.02}$ & -$1.46^{0.02}$ & -$1.46^{0.02}$ & -$1.45^{0.02}$ & -$1.34^{0.03}$ & -$1.32^{0.03}$ & -$1.32^{0.02}$ & -$1.36^{0.03}$\\
\object{$\beta$\,Leo} & $2.30^{0.02}$ & $2.22^{0.02}$ & $2.14^{0.02}$ & $2.08^{0.02}$ & $2.06^{0.02}$ & $2.02^{0.01}$ & $1.99^{0.09}$ & $1.86^{0.09}$ & $1.86^{0.09}$\\
\object{$\alpha$\,PsA} & $1.31^{0.02}$ & $1.25^{0.02}$ & $1.16^{0.02}$ & $1.10^{0.02}$ & $1.08^{0.02}$ & $1.06^{0.05}$ & $1.05^{0.06}$ & $0.99^{0.03}$ & $1.01^{0.07}$\\
\object{$\alpha$\,Cen\,A} & $0.92^{0.02}$ & $0.70^{0.02}$ & $-0.01^{0.02}$ &  &  & -$1.16^{0.02}$ & -$1.39^{0.09}$ & -$1.50^{0.02}$ & -$1.55^{0.09}$\\
{\it Sun}$^{(e)}$ & -$25.98^{0.02}$ & -$26.12^{0.02}$ & -$26.75^{0.02}$ & -$27.12^{0.02}$ & -$27.48^{0.02}$ & -$27.86^{0.02}$ & -$28.20^{0.02}$ & -$28.22^{0.02}$ & \\
\object{$\tau$\,Cet} & $4.43^{0.02}$ & $4.22^{0.02}$ & $3.50^{0.02}$ & $2.88^{0.01}$ & $2.41^{0.01}$ & $2.11^{0.01}$ & $1.73^{0.01}$ & $1.66^{0.01}$ & $1.64^{0.01}$\\
\object{GJ\,166\,A} & $5.69^{0.02}$ & $5.25^{0.02}$ & $4.43^{0.02}$ & $3.72^{0.01}$ & $3.27^{0.01}$ & $2.91^{0.03}$ & $2.46^{0.01}$ & $2.39^{0.02}$ & $2.30^{0.02}$\\
\object{$\alpha$\,Cen\,B} & $2.86^{0.02}$ & $2.21^{0.02}$ & $1.33^{0.02}$ &  &  & -$0.01^{0.02}$ & -$0.49^{0.09}$ & -$0.60^{0.02}$ & -$0.63^{0.09}$\\
\object{$\epsilon$\,Eri} & $5.19^{0.02}$ & $4.61^{0.02}$ & $3.73^{0.02}$ & $3.01^{0.02}$ & $2.54^{0.02}$ & $2.23^{0.03}$ & $1.75^{0.03}$ & $1.67^{0.01}$ & $1.60^{0.05}$\\
\object{GJ\,105\,A} & $7.58^{0.02}$ & $6.81^{0.02}$ & $5.83^{0.02}$ & $4.99^{0.02}$ & $4.46^{0.02}$ & $4.07^{0.03}$ & $3.52^{0.03}$ & $3.45^{0.03}$ & $3.43^{0.03}$\\
\object{GJ\,570\,A} & $7.88^{0.02}$ & $6.82^{0.02}$ & $5.71^{0.02}$ & $4.72^{0.02}$ & $4.18^{0.02}$ & $3.82^{0.02}$ & $3.27^{0.02}$ & $3.15^{0.02}$ & $3.11^{0.02}$\\
\object{$\epsilon$\,Ind\,A} & $6.74^{0.02}$ & $5.75^{0.02}$ & $4.69^{0.02}$ & $3.81^{0.02}$ & $3.25^{0.02}$ & $2.83^{0.02}$ & $2.30^{0.02}$ & $2.18^{0.02}$ & $2.12^{0.02}$\\
\object{GJ\,380} & $9.23^{0.02}$ & $7.94^{0.02}$ & $6.59^{0.02}$ & $5.36^{0.02}$ & $4.56^{0.02}$ & $3.98^{0.03}$ & $3.32^{0.03}$ & $3.19^{0.03}$ & $3.11^{0.03}$\\
\object{GJ\,191} & $11.64^{0.02}$ & $10.40^{0.02}$ & $8.86^{0.02}$ &  &  & $5.77^{0.02}$ & $5.27^{0.02}$ & $5.05^{0.02}$ & $4.86^{0.02}$\\
\object{GJ\,887} & $9.99^{0.02}$ & $8.83^{0.02}$ & $7.35^{0.02}$ &  &  & $4.20^{0.02}$ & $3.60^{0.02}$ & $3.36^{0.02}$ & $3.20^{0.02}$\\
\object{GJ\,205} & $10.63^{0.02}$ & $9.44^{0.02}$ & $7.97^{0.02}$ & $6.53^{0.02}$ & $5.39^{0.02}$ & $4.77^{0.02}$ & $4.06^{0.02}$ & $3.86^{0.02}$ & $3.83^{0.02}$\\
\object{GJ\,15\,A} & $10.88^{0.02}$ & $9.63^{0.02}$ & $8.07^{0.02}$ & $6.72^{0.02}$ & $5.53^{0.02}$ & $4.86^{0.03}$ & $4.25^{0.03}$ & $4.02^{0.02}$ & $3.87^{0.03}$\\
\object{GJ\,411} & $10.13^{0.02}$ & $9.00^{0.02}$ & $7.49^{0.02}$ & $5.98^{0.02}$ & $4.76^{0.02}$ & $4.13^{0.03}$ & $3.56^{0.03}$ & $3.35^{0.03}$ & $3.20^{0.03}$\\
\object{GJ\,699} & $12.57^{0.02}$ & $11.28^{0.02}$ & $9.54^{0.02}$ & $7.71^{0.02}$ & $6.10^{0.02}$ & $5.30^{0.02}$ & $4.77^{0.02}$ & $4.52^{0.02}$ & $4.18^{0.02}$\\
\object{{\it Proxima}} & $14.56^{0.02}$ & $13.02^{0.02}$ & $11.05^{0.02}$ & $8.68^{0.02}$ & $6.42^{0.02}$ & $5.33^{0.02}$ & $4.73^{0.02}$ & $4.36^{0.03}$ & $4.04^{0.02}$\\
\hline
\noalign{\smallskip}
\object{$\gamma$\,Gem} & $1.97^{0.02}$ & $1.92^{0.02}$ & $1.92^{0.02}$ & $1.86^{0.02}$ & $1.87^{0.02}$ & $1.87^{0.02}$ & $1.83^{0.02}$ & $1.85^{0.03}$ & $1.87^{0.09}$\\
\object{$\alpha$\,CMi\,A} & $0.82^{0.02}$ & $0.79^{0.02}$ & $0.37^{0.02}$ & -$0.05^{0.02}$ & -$0.28^{0.02}$ & -$0.40^{0.03}$ & -$0.60^{0.03}$ & -$0.65^{0.03}$ & -$0.68^{0.03}$\\
\object{$\eta$\,Boo} & $3.46^{0.02}$ & $3.26^{0.02}$ & $2.68^{0.02}$ & $2.24^{0.02}$ & $1.95^{0.02}$ & $1.67^{0.02}$ & $1.39^{0.01}$ & $1.35^{0.01}$ & $1.30^{0.02}$\\
\object{$\zeta$\,Her\,A} & $3.67^{0.02}$ & $3.46^{0.02}$ & $2.81^{0.02}$ & $2.30^{0.02}$ & $1.98^{0.02}$ & $1.77^{0.01}$ & $1.42^{0.01}$ & $1.38^{0.01}$ & $1.30^{0.06}$\\
\object{$\mu$\,Her} & $4.57^{0.02}$ & $4.17^{0.02}$ & $3.42^{0.02}$ & $2.89^{0.02}$ & $2.51^{0.02}$ & $2.13^{0.01}$ & $1.80^{0.01}$ & $1.74^{0.01}$ & $1.72^{0.02}$\\
\object{$\beta$\,Aql} & $5.07^{0.02}$ & $4.58^{0.02}$ & $3.72^{0.02}$ & $3.06^{0.02}$ & $2.57^{0.02}$ & $2.19^{0.02}$ & $1.70^{0.01}$ & $1.65^{0.02}$ & $1.61^{0.02}$\\
\object{$\eta$\,Cep} & $4.96^{0.02}$ & $4.35^{0.02}$ & $3.43^{0.02}$ & $2.76^{0.02}$ & $2.27^{0.02}$ & $1.80^{0.05}$ &  & $1.22^{0.02}$ & $1.17^{0.09}$\\
\object{$\delta$\,Eri} & $5.13^{0.02}$ & $4.46^{0.02}$ & $3.54^{0.02}$ & $2.82^{0.02}$ & $2.32^{0.02}$ & $1.95^{0.01}$ & $1.49^{0.02}$ & $1.40^{0.01}$ & $1.36^{0.02}$\\
\hline
\end{tabular}
\begin{itemize}{}{}
\item References:
\item[$(a)$] Morel et al.~(\cite{morel78}), Mermilliod~(\cite{mermilliod86}).
\item[$(b)$] Morel et al.~(\cite{morel78}), Hoffleit \& Warren~(\cite{hoffleit91}), Perryman et al.~(\cite{hip}).
\item[$(c)$] Morel et al.~(\cite{morel78}), Bessell et al.~(\cite{bessell98}), Ducati et al.~(\cite{ducati02}).
\item[$(d)$] Glass~(\cite{glass74}), Glass~(\cite{glass75}), Mould \& Hyland~(\cite{mould76}),
Morel et al.~(\cite{morel78}), Leggett~(\cite{leggett92}), Ducati et al.~(\cite{ducati02}),
Kidger \& Mart\'in-Luis~(\cite{kidger03}), the 2MASS catalogue (Cutri et al.~\cite{cutri03}).
\item[$(e)$] We refered to Colina et al.~(\cite{colina96}) for the apparent magnitudes of the Sun.
\end{itemize}
\end{table*}

\section{Surface brightness relations \label{section_relations}}

\subsection{Fitting procedure}

For each angular diameter measurement $\theta_{\rm LD}$, and based on the observed
apparent magnitudes $m_\lambda$, we have computed the surface brightness $F_\lambda$
in all bands, using the definition of Eq.~(\ref{f_eq}).
The resulting $F_\lambda$ values were then fitted relative to the colors $(C_0 - C_1)$,
using a linear model. This fit was achieved using a classical $\chi^2$
minimization taking into account the errors in both the colors and
$F_\lambda$. The minimized quantity, using the slope $a$ and zero point $b$ as variables, is
the reduced $\chi^2$ expression:
\begin{equation}
\chi^2_{\rm red} (a,b) = \frac{1}{N-2} \sum_{i=1}^{N}
\frac{\left[\left(F_0\right)_i - a\,\left(C_0-C_1\right)_i - b\right]^2}{(\sigma_{F\,i})^2 + a^2\,(\sigma_{C\,i})^2}
\end{equation}
where we have:
\begin{itemize}
\item $N$ the total number of measurements in our sample,
\item $(F_0)_i$ the surface brightness of star $i$ in band $C_0$,
\item $(C_0-C_1)_i$ the color of the star of index $i$ computed between bands $C_0$ and $C_1$,
\item $\sigma_{C\,i}$ the 1\,$\sigma$ error bar in the chosen color of star $i$,
\item $\sigma_{F\,i}$ the 1\,$\sigma$ error bar in the surface brightness $F_0$.
\end{itemize}
The  1\,$\sigma$ errors $\sigma_a$ and $\sigma_b$ are subsequently
estimated from the best fit values $a$ and $b$ by solving numerically the expression:
\begin{equation}
\chi^2_{\rm red}(a+\sigma_a, b+\sigma_b) = \chi^2_{\rm red}(a, b) + 1.
\end{equation}
The solutions of this equation correspond to an elliptic contour, due to the correlation
between the $a$ and $b$ variables. It has to be projected on the $a$ and $b$ axis
to give the errors. The residuals $\Delta F_i = F_i - F_{\rm model}$
are used to estimate the intrinsic dispersion $\sigma_{\rm int}(F)$ of the surface brightness
relation from:
\begin{equation}\label{sigint_expression}
\sigma_{\rm int}^2(F) = \frac{1}{N}\sum_{i=1}^{N}\left[\left(\Delta F_i\right)^2 - (\sigma_{F\,i})^2\right]
\end{equation}
This process gives a total number of 72 $(a,b)$ best fit pairs, with their associated
errors $(\sigma_a, \sigma_b)$, and the intrinsic dispersion $\sigma_{\rm int}$
of the data around the best fit model.

From there, we can invert these relations easily to obtain their angular diameter
counterparts:
\begin{equation}
\log \theta_{\rm LD} = c\,(C_0 - C_1) + d - 0.2\,C_0
\end{equation}
The slopes and zero points are computed from the $(a,b)$ pairs through:
\begin{equation}
c = -2\,a,\ \ \sigma_c = 2\,\sigma_a
\end{equation}
\begin{equation}
d = 2\,(4.2207 - b),\ \ \sigma_d = 2\,\sigma_b
\end{equation}
and the intrinsic dispersions $\sigma_{\rm int}(\log \theta_{\rm LD})$ are given by:
\begin{equation}
\sigma_{\rm int}(\log \theta_{\rm LD}) = 2\,\sigma_{\rm int}(F_\lambda)
\end{equation}

The same method was used for the fits using the effective temperature,
except that no error bar was considered on the $T_{\rm eff}$ values from
the literature (equal weighting), and we used a second degree polynomial
model instead of a linear one. We minimized numerically the following
$\chi^2_{\rm red}$ expression using $a$, $b$, $c$ as variables:
\begin{equation}
\chi^2_{\rm red}(a,b,c) = \frac{1}{N-3} \sum_{i=1}^{N}
\frac{\left[F_i - F_{\rm model}(T_{\rm eff})_i \right]^2}{\sigma_F^2}
\end{equation}
where
\begin{equation}
F_{\rm model}(T_{\rm eff})_i = a\,\left(\log T_{\rm eff}\right)_i^2 + b \left(\log T_{\rm eff}\right)_i + c.
\end{equation}
The errors in each of the $a$, $b$ and $c$ coefficients were not computed,
as the correlations existing between these coefficients make it very difficult to determine
them accurately.
This is justified by the fact that the astrophysical dispersion of the measurements is largely
dominant over the 1\,$\sigma$ fitting errors of the model, and the systematic errors in these coefficients
can thus be considered negligible. The inversion of the resulting $T_{\rm eff}$ based
relations is straightforward. With an expression of the form:
\begin{equation}
\log \theta_{\rm LD} = d (\log T_{\rm eff})^2 + e (\log T_{\rm eff}) + f - 0.2 C_\lambda
\end{equation}
we have by definition:
\begin{equation}
d = -2\,a,\ \ e = -2\,b
\end{equation}
\begin{equation}
f =  2\,(4.2207 - c)
\end{equation}
As in the previous case based on colors, the intrinsic dispersions
$\sigma_{\rm int}(\log \theta_{\rm LD})$ of the angular diameter relations are given by:
\begin{equation}
\sigma_{\rm int}(\log \theta_{\rm LD}) = 2\,\sigma_{\rm int}(F)
\end{equation}

In some cases, we could derive only upper limits of the intrinsic
dispersion $\sigma_{\rm int}$, as it was found to be smaller than
the average error of the measurements (in such cases,
Eq.~\ref{sigint_expression} gives a negative value for $\sigma_{\rm int}$).
For these relations, such as $\log \theta_{\rm LD}(T_{\rm eff}, L)$,
we conclude that the intrinsic dispersion
is undetectable at our level of sensitivity.

\subsection{Angular diameter relations based on colors\label{color_rel_sect}}

\begin{figure}[t]
\centering
\includegraphics[bb=0 0 360 288, width=8.5cm]{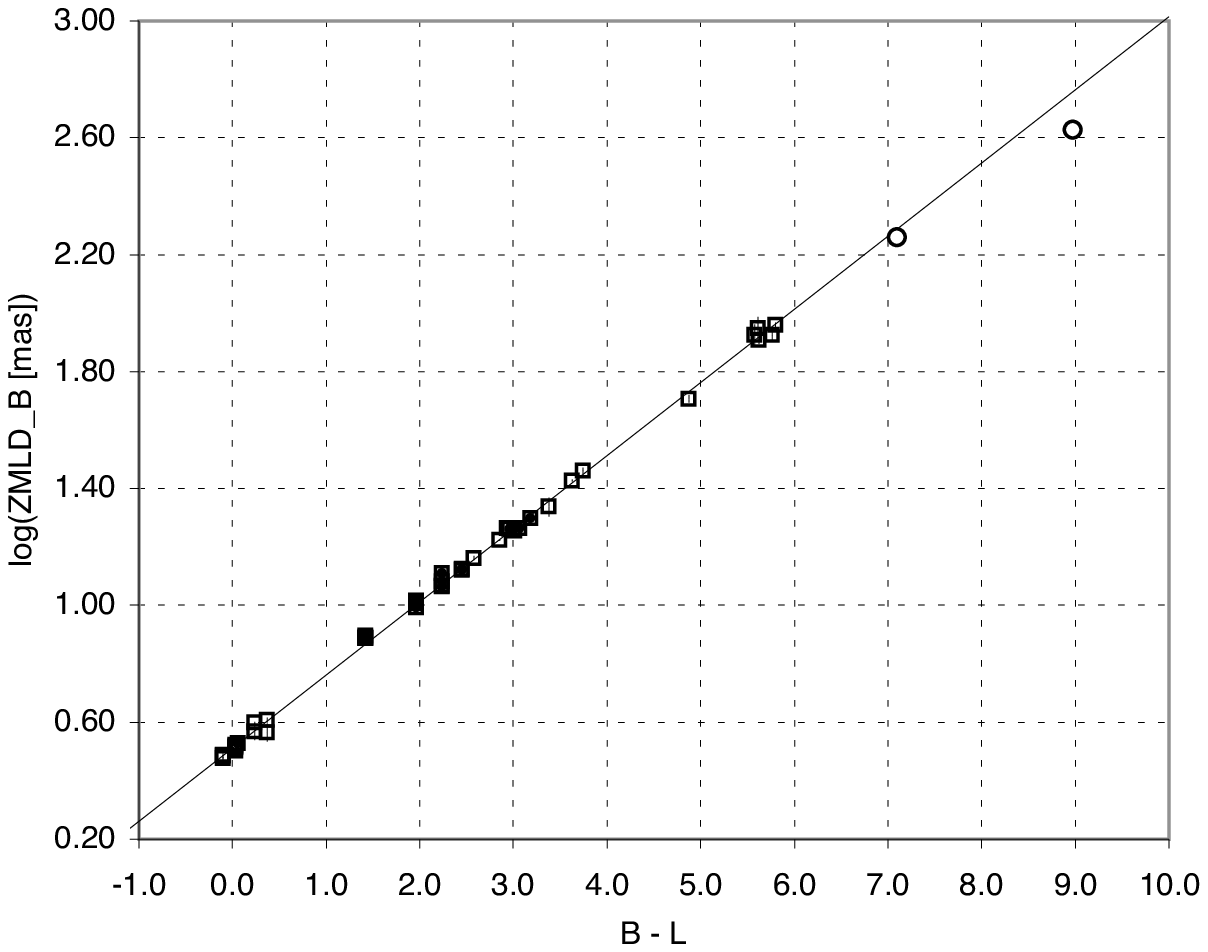}
\includegraphics[bb=0 0 360 144, width=8.5cm]{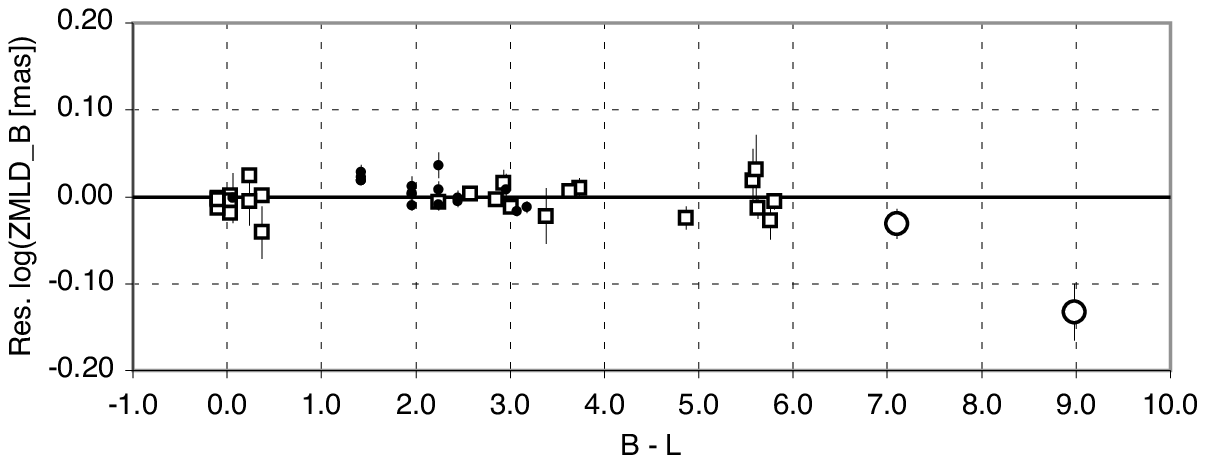}
\caption{Linear fit of the surface brightness relation $\log {\rm ZMLD}_B (B-L)$ (upper part),
and the corresponding residuals (lower part). The intrinsic dispersion in the relation is
$\pm 0.004$ on $\log {\rm ZMLD}$, equivalent to a systematic error of less than 1\,\% in the
predicted angular diameters. The open circles designate GJ\,699 and {\it Proxima},
which were excluded from the fitting procedure.}
\label{Res_ZMLD_B_Color}
\end{figure}

The SB relations for $UBVRIJHKL$ colors are listed in Table~\ref{Angdiam_table_UBVRIJHKL}.
They take the form:
\begin{equation}
\log \theta_{\rm LD}(C_0, C_1) = {c_\lambda (C_0 - C_1)+d_\lambda - 0.2\,C_0}
\end{equation}
where $C_0$ and $C_1$ are any two distinct colors of the Johnson system.
In many cases, the dependence of the zero magnitude limb darkened angular diameter (ZMLD),
defined for $C_0 = 0$, as a function of the color is not linear in reality.
Thus, the linear model that we fit does not represent the observations well.
In this case, we have added a note "{\it nl}" after the obtained residual dispersion.
The non-linear relations should preferably not be used for predictions, though the
stated dispersions include the non-linearity.

In theory, there should be a perfect diagonal symmetry between the dispersions listed in
Table~\ref{Angdiam_table_UBVRIJHKL}. In reality, the symmetry is only approximate, because
$C_0$ and $C_1$ are not symmetric in the expression of $\log \theta_{\rm LD}(C_0, C_1)$.
Therefore, an increased dispersion of the apparent magnitudes in one band $C_1$ will be
reflected preferentially in the $\theta_{\rm LD}(C_0, C_1)$ dispersion, rather than in
$\theta_{\rm LD}(C_1, C_0)$.
For this reason, we provide both versions in Table~\ref{Angdiam_table_UBVRIJHKL},
including the quasi-symmetric pairs. The best relations based on the $K$ band
(showing residual dispersions below 1\% on the angular diameter $\theta_{\rm LD}$)
are the following:
\begin{equation}
\log \theta_{\rm LD} = 0.0535\,(B-K) + 0.5159 - 0.2\,K
\end{equation}
\begin{equation}
\log \theta_{\rm LD} = 0.0755\,(V-K) + 0.5170 - 0.2\,K.
\end{equation}
and the best relations for the $L$ band are:
\begin{equation}
\log \theta_{\rm LD} = 0.0412\,(U-L) + 0.5167 - 0.2\,L
\end{equation}
\begin{equation}
\log \theta_{\rm LD} = 0.0502\,(B-L) + 0.5133 - 0.2\,L
\end{equation}
\begin{equation}
\log \theta_{\rm LD} = 0.0701\,(V-L) + 0.5139 - 0.2\,L
\end{equation}
\begin{equation}
\log \theta_{\rm LD} = 0.1075\,(R-L) + 0.5128 - 0.2\,L.
\end{equation}
These expressions are valid at least over the range of colors defined by our
sample (Tables~\ref{angdiams_table_V} and \ref{angdiams_table_IV}).
In terms of spectral types, the angular diameter predictions can be considered
reliable between A0 and M2 for dwarfs, and between A0 and K0 for subgiants.
There are indications (Fig.~\ref{Res_ZMLD_B_Color}) that the infrared relations are valid down to the
spectral type M4V of GJ\,699, but show some discrepancy for the M5.5V star {\it Proxima}.
The established relations are likely to be valid also for subgiants of spectral types
later than K0IV, but this cannot be verified from our sample.
It should be stressed that they are applicable only to single stars, and the presence
of a non-resolved stellar companion contributing a significant fraction of the measured
flux will bias the predicted angular diameters. As more than half of the Main Sequence
stars are binary or multiple stars, care should be taken in the application of these relations.

\begin{figure*}[t]
\centering
\includegraphics[bb=0 0 500 250, width=17cm]{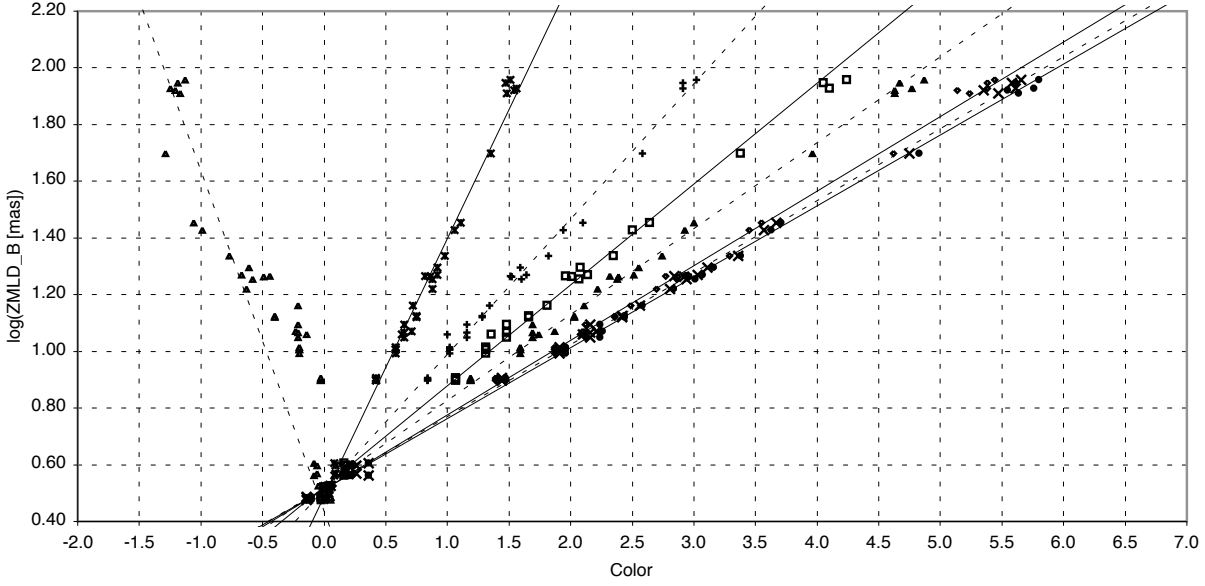}
\caption{Johnson $B$ band ZMLD$_B$ relations as a function of color.
The errors bars have been omitted for clarity, and the fitted models are
represented alternatively as solid and dashed lines. From left to right,
using the colors: $(B-U)$, $(B-V)$, $(B-R)$, $(B-I)$,
$(B-J)$, $(B-H)$, $(B-K)$, $(B-L)$. A clear non-linearity is visible on
the $(B-U)$ based relation.}
\label{ZMLD_B_Color}
\end{figure*}

\begin{table*}
\caption{Surface brightness relations using $UBVRIJHKL$ based colors to obtain
the limb darkened angular diameter $\theta_{\rm LD}$ (in mas) as a function of the magnitude
and color of the star through:
$\log \theta_{\rm LD}(C_0, C_1) = {c_\lambda (C_0 - C_1)+d_\lambda - 0.2\,C_0}$.
The residual dispersions are given in percents of the LD angular diameter.
 The 1\,$\sigma$ errors in each coefficient are given in superscript,
{\it multiplied by 1000} to reduce the length of each line, i.e. $0.5822^{3.2} $ stands for
$0.5822 \pm 0.0032$. When the data depart significantly
from our linear fit and present a detectable non-linearity, the dispersion is mentioned in {\it italic} characters,
and we have added the note "{\it nl}". The dispersions smaller than 5\% are mentioned in bold characters:
they mark the relations that are the most suitable for predicting stellar angular sizes.}
\label{Angdiam_table_UBVRIJHKL}
\begin{tabular}{lrrrrrrrrr}
\hline

$C_0\ \downarrow$ & $C_1\rightarrow U$ & $B$ & $V$ & $R$ & $I$ & $J$ & $H$ & $K$ & $L$ \\
\hline \noalign{\smallskip}
$c_U$ & $$ & $1.2701^{11.8} $ & $0.5822^{3.2} $ & $0.3925^{1.7} $ & $0.3178^{1.1} $ & $0.2805^{1.0} $ & $0.2509^{0.8} $ & $0.2437^{0.8} $ & $0.2407^{0.8} $ \\
$d_U$ & $$ & $ 0.6607^{7.9} $ & $0.5532^{4.5} $ & $0.5393^{3.5} $ & $0.5384^{3.1} $ & $0.5351^{3.0} $ & $0.5206^{2.9} $ & $0.5217^{2.7} $ & $0.5184^{2.9} $ \\
$\sigma_U$ & $$ & $ \it57.8\%\ {\it nl}$ & $ \it17.9\%\ {\it nl}$ & $ \it10.6\%\ {\it nl}$ & $ \it7.88\%\ {\it nl}$ & $ \it5.05\%\ {\it nl}$ & $ \bf3.06\%$ & $ \bf2.76\%$ & $ \bf1.92\%$\\
\noalign{\smallskip} \hline \noalign{\smallskip}
$c_B$ & -$1.0923^{10.4} $ & $$ & $0.9095^{6.9} $ & $0.4771^{2.2} $ & $0.3557^{1.4} $ & $0.3029^{1.0} $ & $0.2630^{0.8} $ & $0.2538^{0.7} $ & $0.2501^{0.9} $ \\
$d_B$ & $ 0.6542^{7.0} $ & $$ & $ 0.4889^{1.1} $ & $0.5116^{0.1} $ & $0.5235^{0.1} $ & $0.5242^{0.1} $ & $0.5134^{1.0} $ & $0.5158^{0.1} $ & $0.5133^{0.1} $ \\
$\sigma_B$ & $ \it59.2\%\ {\it nl}$ & $$ & $ \it9.17\%\ {\it nl}$ & $ \it4.75\%\ {\it nl}$ & $ \it4.98\%\ {\it nl}$ & $ \bf3.07\%$ & $ \bf1.89\%$ & $ \bf \le1.00\%$ & $ \bf \le1.00\%$\\
\noalign{\smallskip} \hline \noalign{\smallskip}
$c_V$ & -$0.3846^{2.3} $ & -$0.7083^{7.4} $ & $$ & $0.7900^{8.4} $ & $0.4550^{2.9} $ & $0.3547^{2.0} $ & $0.2893^{1.6} $ & $0.2753^{1.4} $ & $0.2694^{1.4} $ \\
$d_V$ & $ 0.5513^{3.4} $ & $0.4889^{5.7} $ & $$ & $ 0.5217^{6.4} $ & $0.5332^{4.0} $ & $0.5310^{3.6} $ & $0.5148^{3.2} $ & $0.5175^{2.9} $ & $0.5146^{3.1} $ \\
$\sigma_V$ & $ \it18.0\%\ {\it nl}$ & $ \it9.18\%\ {\it nl}$ & $$ & $ \it8.53\%\ {\it nl}$ & $ \it8.30\%\ {\it nl}$ & $ \bf4.75\%$ & $ \bf1.85\%$ & $ \bf1.01\%$ & $ \bf \le1.00\%$\\
\noalign{\smallskip} \hline \noalign{\smallskip}
$c_R$ & -$0.1966^{1.0} $ & -$0.2792^{1.7} $ & -$0.5842^{5.1} $ & $$ & $0.7404^{6.4} $ & $0.4647^{3.2} $ & $0.3405^{2.5} $ & $0.3158^{1.9} $ & $0.3041^{1.8} $ \\
$d_R$ & $ 0.5351^{2.4} $ & $0.5109^{3.0} $ & $0.5251^{4.7} $ & $$ & $ 0.5570^{5.2} $ & $0.5392^{4.0} $ & $0.5119^{3.8} $ & $0.5158^{3.1} $ & $0.5144^{3.3} $ \\
$\sigma_R$ & $ \it11.1\%\ {\it nl}$ & $ \it5.02\%\ {\it nl}$ & $ \it8.63\%\ {\it nl}$ & $$ & $ \it14.5\%\ {\it nl}$ & $ \it8.43\%\ {\it nl}$ & $ \bf2.88\%$ & $ \bf2.05\%$ & $ \bf2.52\%$\\
\noalign{\smallskip} \hline \noalign{\smallskip}
$c_I$ & -$0.1210^{0.7} $ & -$0.1587^{1.1} $ & -$0.2609^{2.0} $ & -$0.5998^{7.6} $ & $$ & $0.9079^{19.8} $ & $0.4318^{6.0} $ & $0.3833^{4.5} $ & $0.3707^{4.6} $ \\
$d_I$ & $ 0.5351^{1.9} $ & $0.5207^{2.3} $ & $0.5296^{2.8} $ & $0.5416^{4.7} $ & $$ & $ 0.5233^{7.6} $ & $0.5089^{4.4} $ & $0.5140^{3.7} $ & $0.5097^{4.0} $ \\
$\sigma_I$ & $ \it8.30\%\ {\it nl}$ & $ \it5.34\%\ {\it nl}$ & $ \it8.73\%\ {\it nl}$ & $ \it12.1\%\ {\it nl}$ & $$ & $ \it10.8\%\ {\it nl}$ & $ \it5.83\%\ {\it nl}$ & $ \bf3.84\%$ & $ \bf3.30\%$\\
\noalign{\smallskip} \hline \noalign{\smallskip}
$c_J$ & -$0.0818^{0.6} $ & -$0.1043^{0.9} $ & -$0.1581^{1.4} $ & -$0.2842^{3.2} $ & -$0.7198^{16.9} $ & $$ & $0.6280^{16.0} $ & $0.5214^{10.6} $ & $0.4840^{9.4} $ \\
$d_J$ & $ 0.5325^{1.9} $ & $0.5216^{2.2} $ & $0.5276^{2.4} $ & $0.5299^{3.2} $ & $0.5197^{6.5} $ & $$ & $ 0.4990^{5.8} $ & $0.5066^{4.9} $ & $0.5078^{4.8} $ \\
$\sigma_J$ & $ \it5.13\%\ {\it nl}$ & $ \bf3.12\%$ & $ \bf4.98\%$ & $ \it7.26\%\ {\it nl}$ & $ \it10.9\%\ {\it nl}$ & $$ & $ \it10.44\%\ {\it nl}$ & $ \it5.86\%\ {\it nl}$ & $ \bf4.08\%$\\
\noalign{\smallskip} \hline \noalign{\smallskip}
$c_H$ & -$0.0513^{0.5} $ & -$0.0625^{0.6} $ & -$0.0892^{0.9} $ & -$0.1376^{1.8} $ & -$0.2290^{4.0} $ & -$0.4312^{12.1} $ & $$ & $1.8747^{58.8} $ & $1.1714^{27.7} $ \\
$d_H$ & $ 0.5189^{1.6} $ & $0.5138^{1.8} $ & $0.5145^{1.9} $ & $0.5138^{2.3} $ & $0.5093^{2.9} $ & $0.5013^{4.4} $ & $$ & $ 0.5352^{7.5} $ & $0.5271^{6.2} $ \\
$\sigma_H$ & $ \bf2.67\%$ & $ \bf1.24\%$ & $ \bf1.12\%$ & $ \bf2.06\%$ & $ \it5.53\%\ {\it nl}$ & $ \it10.31\%\ {\it nl}$ & $$ & $ \it17.2\%\ {\it nl}$ & $ \it13.0\%\ {\it nl}$\\
\noalign{\smallskip} \hline \noalign{\smallskip}
$c_K$ & -$0.0440^{0.4} $ & -$0.0535^{0.6} $ & -$0.0755^{0.8} $ & -$0.1144^{1.4} $ & -$0.1805^{2.8} $ & -$0.3192^{7.4} $ & -$1.7040^{55.1} $ & $$ & $3.0857^{33.9} $ \\
$d_K$ & $ 0.5202^{1.6} $ & $0.5159^{1.6} $ & $0.5170^{1.7} $ & $0.5168^{1.9} $ & $0.5149^{2.3} $ & $0.5089^{3.4} $ & $0.5336^{7.0} $ & $$ & $ 0.6258^{3.6} $ \\
$\sigma_K$ & $ \bf2.58\%$ & $ \bf \le1.00\%$ & $ \bf \le1.00\%$ & $ \bf1.60\%$ & $ \bf3.67\%$ & $ \it5.87\%\ {\it nl}$ & $ \it17.6\%\ {\it nl}$ & $$ & $ \it26.9\%\ {\it nl}$\\
\noalign{\smallskip} \hline \noalign{\smallskip}
$c_L$ & -$0.0412^{0.5} $ & -$0.0502^{0.6} $ & -$0.0701^{0.8} $ & -$0.1075^{1.4} $ & -$0.1696^{2.9} $ & -$0.2826^{6.4} $ & -$0.9843^{24.1} $ & -$2.8950^{115.7} $ & $$\\
$d_L$ & $ 0.5167^{1.7} $ & $0.5133^{1.8} $ & $0.5139^{1.9} $ & $0.5128^{2.1} $ & $0.5101^{2.5} $ & $0.5081^{3.3} $ & $0.5245^{5.4} $ & $0.5309^{12.4} $ & $$ \\
$\sigma_L$ & $ \bf \le1.00\%$ & $ \bf \le1.00\%$ & $ \bf \le1.00\%$ & $ \bf \le1.00\%$ & $ \bf2.43\%$ & $ \bf3.61\%$ & $ \it13.1\%\ {\it nl}$ & $ \it26.3\%\ {\it nl}$ & $$\\
\noalign{\smallskip} \hline
\end{tabular}
\end{table*}

\subsection{Angular diameter relations based on effective temperatures \label{Teff_relations}}

\begin{figure}[t]
\centering
\includegraphics[bb=0 0 360 288, width=8.5cm]{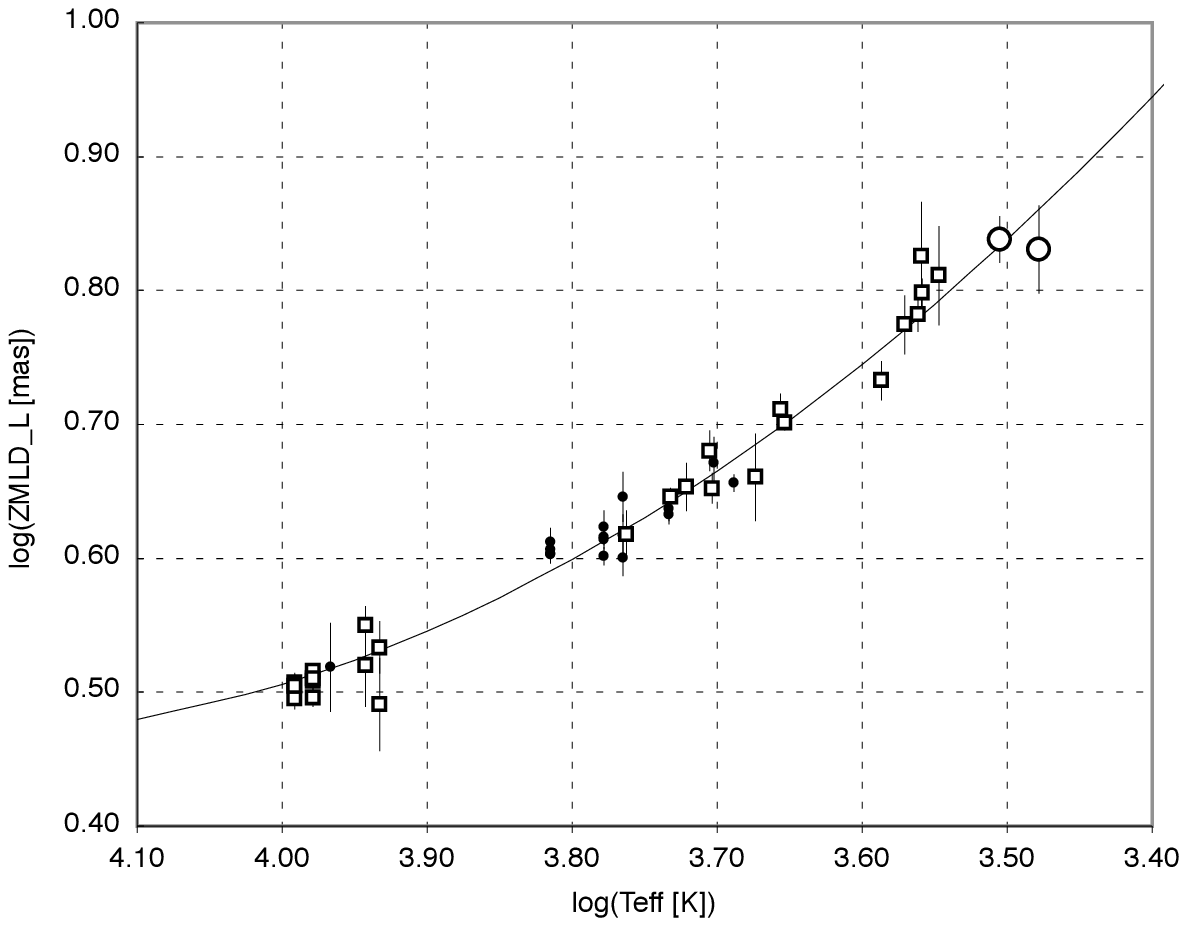}
\includegraphics[bb=0 0 360 144, width=8.5cm]{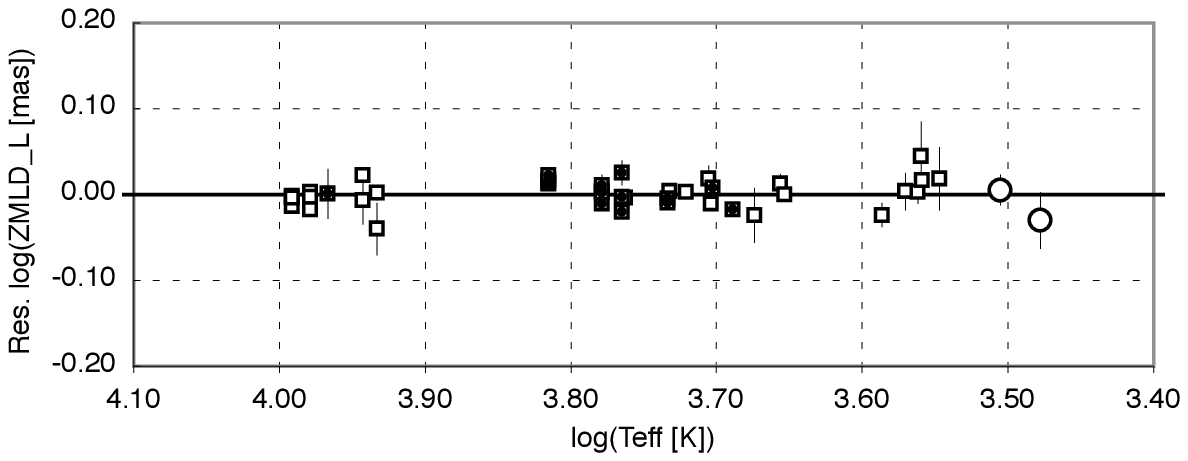}
\caption{Second degree polynomial fit of ZMLD$_L(\log T_{\rm eff})$ (upper part)
and the corresponding residuals (lower part).
The coefficients are given in Table~\ref{Angdiam_table_Teff}, and correspond
to a relation of the form $\log {\rm ZMLD}_L = d (\log T_{\rm eff})^2 + e (\log T_{\rm eff}) + f$.
The open circles designate GJ\,699 and {\it Proxima}, which were excluded
from the fit (see Sect.~\ref{selection}), though they are consistent
with the model within their error bars.}
\label{ZMLD_L_Teff}
\end{figure}

Table~\ref{Angdiam_table_Teff} gives the best fit model coefficients for the
relations $\theta_{\rm LD} (T_{\rm eff}, C_\lambda)$, defined as:
\begin{equation}\label{LD_Teff_eq}
\log \theta_{\rm LD} = d\, (\log T_{\rm eff})^2 + e\, (\log T_{\rm eff}) + f - 0.2\,C_\lambda
\end{equation}
The smallest residuals are obtained for the relations based on $T_{\rm eff}$ and the $K$ or $L$
magnitudes, with an upper limit on the $1\,\sigma$ dispersion of 1.0\%
(the true dispersion is undetectable from our data):
\begin{displaymath}
\log \theta_{\rm LD} = 0.8470 \,x^2 -7.0790 \,x + 15.2731 - 0.2\,K
\end{displaymath}
\begin{displaymath}
\log \theta_{\rm LD} = 0.6662 \,x^2 -5.6609 \,x + 12.4902 - 0.2\,L
\end{displaymath}
where $x = \log T_{\rm eff}$.
The range of validity of the $T_{\rm eff}$ based relations is 3600--10000\,K for dwarfs, and
4900--9500\,K for subgiants. As shown in Fig.~\ref{ZMLD_L_Teff}, there are indications
that the infrared relations are valid for dwarfs with $T_{\rm eff}$ down to $\sim$3000\,K.

\begin{figure*}[t]
\centering
\includegraphics[bb=0 0 500 250, width=17cm]{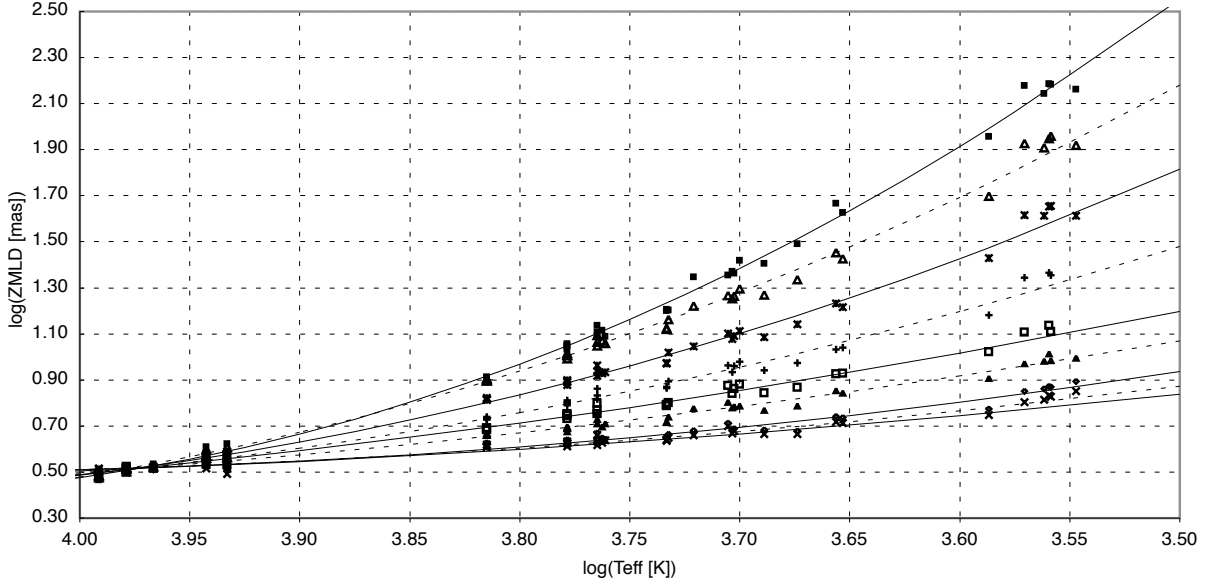}
\caption{ZMLD$_\lambda$ relations as a function of the effective temperature.
The error bars have been omitted for clarity, and the fitted models are
represented alternatively as solid and dashed lines. From top to bottom,
using the zero magnitude reference colors $U$, $B$, $V$, $R$, $I$,
$J$, $H$, $K$ and $L$.}
\label{ZMLD_Teff}
\end{figure*}

\begin{table}
\caption{SB relations using the magnitude $C_\lambda$ and the effective temperature $T_{\rm eff}$
of the star to obtain the limb darkened angular diameter $\theta_{\rm LD}$ (in mas):
$\log \theta_{\rm LD}(T_{\rm eff}, C_{\lambda}) = {d (\log T_{\rm eff})^2+e \log T_{\rm eff} + f - 0.2\,C_{\lambda}}$.
The $1\,\sigma$ residual dispersions are given in percents of the LD angular diameter.}
\label{Angdiam_table_Teff}
\begin{tabular}{lcccc}
\hline
$C_\lambda$ & $\sigma$ & $d$ & $e$ & $f$ \\
\noalign{\smallskip} \hline \noalign{\smallskip}
$U$ & $5.93\%$ & $5.6391$ & $-46.4505$ & $96.0513$ \\
$B$ & $6.33\%$ & $3.6753$ & $-30.9671$ & $65.5421$ \\
$V$ & $5.90\%$ & $3.0415$ & $-25.4696$ & $53.7010$ \\
$R$ & $4.76\%$ & $2.1394$ & $-18.0221$ & $38.3497$ \\
$I$ & $2.28\%$ & $0.9847$ & $-8.7985$ & $19.9281$ \\
$J$ & $1.19\%$ & $0.9598$ & $-8.3451$ & $18.5204$ \\
$H$ & $ 1.38\%$ & $1.1684$ & $-9.6156$ & $20.2779$ \\
$K$ & $ \le1.00\%$ & $0.8470$ & $-7.0790$ & $15.2731$ \\
$L$ & $ \le1.00\%$ & $0.6662$ & $-5.6609$ & $12.4902$ \\
\noalign{\smallskip} \hline
\end{tabular}
\end{table}

\subsection{$T_{\rm eff}(\theta_{\rm LD}, m_\lambda)$ relations}

By inverting the relations established in Sect.~\ref{Teff_relations}, it is possible
to predict the effective temperature of the observed stars based on their angular
diameter and broadband magnitude in a single band. As in the previous sections,
we assume zero interstellar extinction, and the relations are valid only for
dereddened magnitudes. The formulation of the $T_{\rm eff}(\theta_{\rm LD}, m_\lambda)$
laws is easily derived analytically. From Eq.~\ref{LD_Teff_eq},
we obtain $\log T_{\rm eff}$ through the expression:
\begin{equation}
\log T_{\rm eff} = \frac{-\sqrt{{4d \log \theta_{\rm LD} + 0.8d\,C_\lambda} + e^2 - 4df } - e}{2d} 
\end{equation}
that can be rewritten as
\begin{equation}
\log T_{\rm eff} = - \sqrt{g \log \theta_{\rm LD} + h\,C_{\lambda} + i} + j
\end{equation}
where
\begin{equation}
g = \frac{1}{d},\ \ \ \ h = \frac{0.2}{d},
\end{equation}
\begin{equation}
i = \frac{e^2}{4d^2}-\frac{f}{d},\ \ \ \ j = \frac{-e}{2\,d}.
\end{equation}
The intrinsic dispersion of the $\log T_{\rm eff}$ relations can be approximated
from the intrinsic dispersion of the $\log \theta_{\rm LD}$ relations, as we have
$\sigma_{\rm int}(\log \theta_{\rm LD}) \ll 1$:
\begin{equation}
\sigma_{\rm int}(\log T_{\rm eff}) = 0.5\,\sqrt{g}\ \sigma_{\rm int}(\log \theta_{\rm LD})
\end{equation}
The corresponding coefficients and dispersions are given in Table~\ref{inv_Teff_relations}.
The $K$ band relation presents the smallest intrinsic dispersion $(\sigma \le 0.60\%)$,
corresponding to a systematic uncertainty of less than 40\,K in the predicted temperature
of a G2V star:
\begin{displaymath}
\log T_{\rm eff} = 4.1788 - \sqrt{1.1806 \log \theta_{\rm LD} + 0.2361\,K -0.5695}
\end{displaymath}
However, we would like to stress that the uncertainty in the measured apparent
magnitudes can easily be dominant, as a $\pm 0.03$ error in $K$ will translate into a
$\pm 1.7\%$ error in $T_{\rm eff}$, nearly three times as large as the intrinsic dispersion.

Considering that photometry at an absolute level of $\pm 0.01$ is not
available for all stars, the $T_{\rm eff}$ predictions from different bands can be averaged,
taking carefully into account the statistical and systematic
errors of each relation used, in order to reach the intrinsic dispersion level.
In addition, such an averaging process should not be done for stars affected
by interstellar or circumstellar extinction, as it will affect differently each
photometric band.

\begin{table}
\caption{$T_{\rm eff}(\theta_{\rm LD}, m_\lambda)$ relations
to obtain the effective temperature:
$\log T_{\rm eff} = - \sqrt{g \log \theta_{\rm LD} + h\,C_{\lambda} + i} + j$.
The $1\,\sigma$ residual dispersions are given in percents of the effective temperature
$T_{\rm eff}$ (expressed in K).}
\label{inv_Teff_relations}
\begin{tabular}{lccccc}
\hline

$C_\lambda$ & $\sigma$ & $g$ & $h$ & $i$ & $j$ \\
\noalign{\smallskip} \hline \noalign{\smallskip}
$U$ & $1.19\%$ & $0.1773$ & $0.0355$ & $-0.0703$ & $4.1186$ \\
$B$ & $1.57\%$ & $0.2721$ & $0.0544$ & $-0.0850$ & $4.2128$ \\
$V$ & $1.61\%$ & $0.3288$ & $0.0658$ & $-0.1249$ & $4.1870$ \\
$R$ & $1.57\%$ & $0.4674$ & $0.0935$ & $-0.1848$ & $4.2120$ \\
$I$ & $1.13\%$ & $1.0155$ & $0.2031$ & $-0.2788$ & $4.4675$ \\
$J$ & $0.60\%$ & $1.0419$ & $0.2084$ & $-0.3975$ & $4.3472$ \\
$H$ & $0.63\%$ & $0.8559$ & $0.1712$ & $-0.4230$ & $4.1149$ \\
$K$ & $\le 0.60\%$ & $1.1806$ & $0.2361$ & $-0.5695$ & $4.1788$ \\
$L$ & $\le 0.60\%$ & $1.5010$ & $0.3002$ & $-0.6977$ & $4.2486$ \\
\noalign{\smallskip} \hline
\end{tabular}
\end{table}

\subsection{Metallicity}

\begin{figure}[t]
\centering
\includegraphics[bb=0 0 360 288, width=8.5cm]{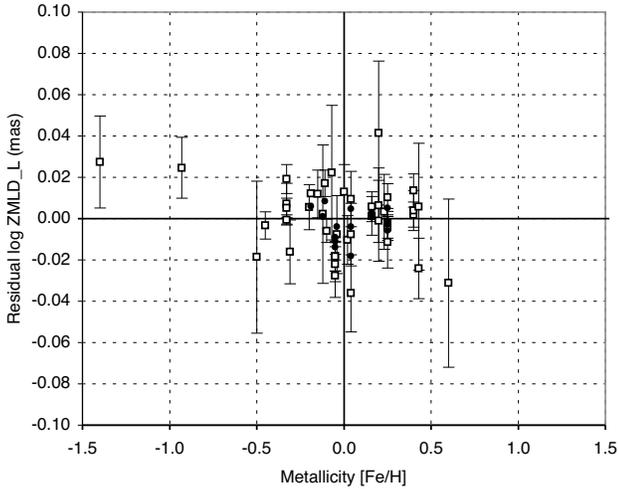}
\caption{Residuals of the fit of ZMLD$_L(L,\,B-L)$ as a function of the metallicity [Fe/H] of
the star. Dwarfs are represented by open squares, and subgiants by solid dots.
No correlation is visible among the stars of our sample.}
\label{Res_metallicity}
\end{figure}

A possible source of natural dispersion of the SB relations is the presence of
deep absorption lines in the spectra of the stars. This effect is stronger for stars that
have a high metal content. However, as shown in Fig.~\ref{Res_metallicity},
there is no clear evidence of a correlation between the residuals of the least
dispersed relation $\theta_{\rm LD}(L,\,B-L)$ and the metallicity [Fe/H]. This is an indication
that our SB relations are valid at least for metallicities between $-0.5$ and $+0.5$~dex,
and probably also for lower values. The two metal-deficient stars
of Fig.~\ref{Res_metallicity} are GJ\,15\,A ([Fe/H]$ = -1.40$ dex)
and GJ\,699 ([Fe/H]$=-0.90$ dex, not included in our fits).
For typical stars of the solar neighborhood our relations are thus always applicable.

\section{Sources of uncertainty in the predicted diameters \label{estim_errors}}

Several observational and astrophysical sources of uncertainty
add up to create the total error in the predicted angular diameters:
\begin{itemize}
\item {\it intrinsic dispersion of the empirical relation:}
as discussed above, the best relations have intrinsic dispersions below 1\%. It should be
stressed that their predictions cannot be averaged to reduce this systematic uncertainty.
However, the predictions from independent colors (such as $V-K$ and $B-L$) can be
averaged to reduce the statistical uncertainty in the predictions due to
the errors in the photometric measurements. In this process, the systematic
uncertainties of each relation {\it cannot} be reduced and have to be carefully
taken into account. This is essential as the calibrations have been obtained for all colors
from the same sample of stars, and the resulting systematic errors are therefore highly correlated.

\item {\it uncertainties in the apparent magnitudes:}
combining the high precision magnitudes available in the visible with the infrared magnitudes
produced by the 2MASS and DENIS surveys should make it possible to retrieve the visible-infrared
color indices with a precision better than $\pm 0.02$\,mag. However, we would like to stress that
the uncertainty in the apparent magnitude measurements can easily be the largest contributor
to the predicted angular diameter errors. The true errors in the photometric measurements
have to be estimated accurately in order to obtain reasonable uncertainties in the
predicted angular sizes.

\item {\it interstellar extinction and circumstellar matter:}
for the sample of nearby stars that was considered for our fits, the interstellar extinction
is negligible: apart from $\gamma$\,Gem at 32\,pc, all stars are located closer than 15\,pc.
However, our SB relations are strictly valid only for extinction-corrected magnitudes.
The uncertainty in the assumed color excess $E(B-V)$ (for instance) will
translate into an additional uncertainty in the dereddened magnitudes.
The presence of a significant amount of circumstellar matter around the star
will also affect its spectral properties, and can be difficult to detect.
\end{itemize}

\section{Comparison between interferometers \label{comparison_section}}

The residuals of the fits of the least dispersed relations (based on infrared colors)
allow us to examine if systematic discrepancies are detectable between the
five interferometers represented in our sample.
For each instrument we have computed the average
residuals of its measurements, and the 1\,$\sigma$ error resulting
from the averaging of their respective errors. The results are presented
in Table~\ref{comparison_interf}.

We observe that the average residuals are below 1.5\,\% in terms of ZMLD
for all instruments. In addition, all the deviations are below 1\,$\sigma$, and
can therefore be fully explained by random statistical dispersion. As a remark,
the agreement between the VLTI/VINCI results and the Mk\,III is remarkable, with
no systematic deviation detectable at a level of a few tenths of a percent. This is
especially encouraging as these two instruments are observing at very different
wavelengths (visible and $K$ band, respectively).

This comparison exercise relies implicitly on the
assumption that the considered ${\rm ZMLD}_0 (C_0-C_1)$ relations
are applicable to each instrument's subsample of stars, down to the precision of each
individual measurement. This may not be the case for all stars, but the agreement
that we observe is a worst case, and the true agreement is in any
case very satisfactory.

\begin{table*}
\caption{Comparison between interferometers. The average residuals of the fits
of ${\rm ZMLD}_B$ and ${\rm ZMLD}_V$ for the $K$ and $L$ based colors
are given together with the corresponding $1\,\sigma$ error bars. All values are
expressed in percents of the ZMLD values. All the residuals are compatible with
zero within their 1\,$\sigma$ error bars.}
\label{comparison_interf}
\begin{tabular}{lccccc}
\hline
Instrument&$N$& $\Delta{\rm ZMLD}_B (B-K)$ & $\Delta {\rm ZMLD}_V (V-K)$ &
$\Delta {\rm ZMLD}_B (B-L)$ & $\Delta {\rm ZMLD}_V (V-L)$ \\
\hline \noalign{\smallskip}
PTI&5& $+0.48\pm0.89\,\% $ & $+0.82\pm0.94\,\% $ & $+0.63\pm0.89\,\% $ & $+0.89\pm0.93$ \\
NII&5& $+1.02\pm1.42\,\% $ & $+1.03\pm1.46\,\% $ & $+1.03\pm1.58\,\% $ & $+1.02\pm1.63$ \\
Mk\,III&11& $-0.11\pm0.56\,\% $ & $-0.12\pm0.57\,\% $ & $-0.06\pm0.64\,\% $ & $-0.11\pm0.66$ \\
NPOI&5& $-1.20\pm1.20\,\% $ & $-1.23\pm1.24\,\% $ & $-1.23\pm1.24\,\% $ & $-1.39\pm1.53$ \\
VLTI/VINCI&16& $-0.05\pm0.49\,\% $ & $-0.07\pm0.51\,\% $ & $-0.07\pm0.51\,\% $ & $-0.11\pm0.68$ \\
\noalign{\smallskip} \hline  
\end{tabular}
\end{table*}

\section{Previous calibrations and other luminosity classes \label{compare_rel}}

Previous calibrations of the SB relations for dwarfs have been derived by
Di Benedetto~(\cite{dibenedetto98}) and Van Belle~(\cite{vanbelle99a}).
These two authors relied on the limited sample of hot dwarfs observed
with the Narrabri intensity interferometer (Hanbury Brown et al.~\cite{hanbury74a};
\cite{hanbury74b}). The agreement of our calibration with the work by
Van Belle~(\cite{vanbelle99a}) is satisfactory within $1\,\sigma$ for
the $(V, V-K)$ relation, but there is a difference of about $2\,\sigma$ in the slope
of the $(B, B-K)$ relation. As the fit obtained by this author is based on
a small range of colors, we attribute this moderate discrepancy to an underestimation of the true
error bar in the slope, even in the restricted quoted range of validity
($-0.6 \le B-K \le +2.0$).

Several calibrations of the SB relations for giants have been proposed in
recent years, thanks to the availability of a number of direct interferometric
measurements of this class of stars.
Van Belle~(\cite{vanbelle99a}) used a sample of 190 giants, complemented by
67 carbon stars and Miras measured with the PTI (Van Belle et al.~\cite{vanbelle99b}),
IOTA (e.g. Dyck et al.~\cite{dyck98}) and lunar occultation observations
(e.g. Ridgway et al.~\cite{ridgway82}) to calibrate the $F_V (V-K)$ relation
of giant and supergiant stars.
Welch~(\cite{welch94}) and Fouqu\'e \& Gieren~(\cite{fouque97})
proposed a calibration of the SB relations of Cepheids based on an
extrapolation of the corresponding relations of giants.
Among the supergiants, Cepheids occupy a particular place.
The observations of these variable stars by interferometry, intended primarily to study their
pulsation, have resulted in the measurement of several of these
objects (Mourard et al.~\cite{mourard97}; Lane et al.~\cite{lane00}; Nordgren et al.~\cite{nordgren00};
Kervella et al.~\cite{kervella01}; Lane et al.~\cite{lane02}; Kervella et al.~\cite{kervella04b}).
Based on these observations, Nordgren et al.~(\cite{nordgren02}) have established dedicated
SB relations for Cepheids, and they find a satisfactory agreement with previous works.
From these studies, it appears that the SB relations found for giants and supergiants are
similar to the ones determined in the present paper for dwarfs and subgiants, especially
their visible-infrared versions. This means qualitatively that any two stars of class I-V
with similar magnitudes in two bands will present approximately the same
angular diameters.

\section{Main Sequence stars as calibrators for long-baseline interferometry \label{calib_interf}}

\subsection{The need for small and nearby calibrators}

Interferometric observations are generally based on interleaved observations of
a scientific target and a calibrator. The angular size of the calibrator is supposed
to be known {\it a priori}, and the observed fringe contrast is used to estimate
the instrumental transfer function (also called system visibility).
The catalogue of calibrators assembled by Cohen et al.~(\cite{cohen99}), and customized to interferometry
by Bord\'e et al.~(\cite{borde02}), consists mainly of K giants with angular diameters
of about 2\,mas. While this size is well adapted to short baseline observations (up to a
few tens of meters in the infrared), these stars are too large angularly to serve as calibrators
for the hectometric baselines of the VLTI, the CHARA array (McAlister et al.~\cite{mcalister00})
or the NPOI (Armstrong et al.~\cite{armstrong98}). In addition, it is foreseen that shorter
wavelengths will be implemented on the VLTI 
than the $K$ band currently accessible with VINCI.
For instance, the AMBER instrument (Petrov et al.~\cite{petrov00})
will allow observations in the $J$ band.
The two-fold increase in angular resolution will naturally require significantly smaller
calibrators than those in the Cohen et al.~(\cite{cohen99}) catalogue.

A fundamental problem with distant stars is that the reddening
corrections are uncertain. This means that it is highly desirable to use
nearby stars as calibrators, located within a few tens of parsecs.
In this respect, giant stars are not well suited due to their large linear dimensions,
but dwarfs and subgiants are ideally suited to
provide small and well-defined calibrators.

Another advantage of Main Sequence stars is that their strong surface gravity
results in a compact atmosphere and a well-defined photosphere.
Their disk appears sharper than that of the giants, for which the precise definition
of the limb darkened disk angular diameter at a level of less than 1\,\%
can be difficult, in particular for the later spectral types.
As an example, a discussion of the M4III giant $\psi$\,Phe can be found in
Wittkowski, Aufdenberg \& Kervella~(\cite{wittkowski04}).

\subsection{Calibration precision vs. brightness}

\begin{figure}[t]
\centering
\includegraphics[bb=0 0 360 288, width=8.5cm]{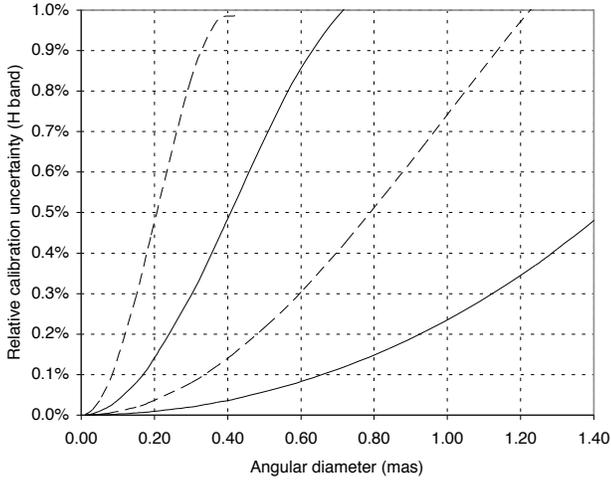}
\caption{Precision achievable in the measurement of the interferometric efficiency as a
function of the angular diameter of the calibrator, predicted using the $(B, B-L)$
SB relation ($\le 1.0\%$ dispersion).
From left to right, the curves refer to baselines of 800, 400, 200 and 100\,m, in the $H$ band.}
\label{Precision_calib}
\end{figure}

It is possible to estimate the maximum angular size of calibrator stars in order to
obtain a given relative precision in the calibration of the interferometric efficiency.
Fig.~\ref{Precision_calib} shows the achievable precision in the interferometric
efficiency using the $(B, B-L)$ relation determined in Sect.~\ref{color_rel_sect}
($\sigma \le 1.0\%$), as a function of the angular diameter of the calibrator star, for
four different baselines (100, 200, 400 and 800\,m), in the $H$ band.
These baselines are representative of the existing or foreseen interferometers
(Keck, PTI, VLTI, CHARA, NPOI and OHANA, sorted by increasing maximum baseline).
The horizontal scale of Fig.~\ref{Precision_calib} can be adapted for other
wavelengths or other baselines by scaling it linearly while maintaining constant
the $B / \lambda$ ratio. 

\begin{figure}[t]
\centering
\includegraphics[bb=0 0 360 288, width=8.5cm]{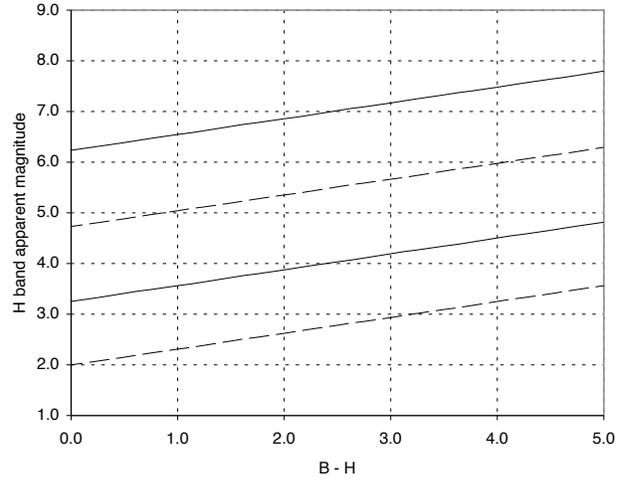}
\caption{Apparent magnitude in the $H$ band of the calibrators suitable for obtaining a
relative precision of 0.5\% in the calibration of the interferometric efficiency,
as a function of the $B-H$ color.
From top to bottom, the curves refer to baselines of 800, 400, 200 and 100\,m,
in the $H$ band.}
\label{Hmag_calib}
\end{figure}

If we now set a limit of 0.5\% on the acceptable systematic uncertainty in the interferometric
efficiency, we can compute the apparent magnitude of the Main Sequence calibrators that
should be used as a function of their color. The result is presented in
Fig.~\ref{Hmag_calib} as a function of the $B-H$ color, for different baseline lengths and
interferometric observations in the $H$ band.
From this figure, it can be concluded that suitable calibrators for extremely long baseline
observations will have to be faint. Let us consider the example of the OHANA interferometer
(original idea proposed by Mariotti et al.~\cite{mariotti96}), whose
longest foreseen baseline is 800\,m. The $H$ band magnitude of the calibrators necessary
to obtain a relative systematic visibility error of 0.5\% will be between $m_H = 6$ and 8,
depending on the spectral type. This is rather faint, even for large aperture light collectors,
but it is feasible with OHANA.

As an alternative, it is possible to build (through
time consuming observations) a secondary network of brighter and larger
calibrators based on the small angular diameter ones, but there will always be a limitation
attached to the fact that calibrators have to be observed in the first lobe of their visibility
function. For OHANA, this sets a hard limit of $\simeq$0.5\,mas on the calibrator angular size,
and even $\le 0.4$\,mas to obtain a visibility of at least 0.3. This corresponds to apparent
magnitudes of $m_H = 5$ to 7 in the $H$ band, one magnitude brighter than the
primary network.

For the longest baseline of the VLTI (200\,m), calibrator magnitudes between $m_H = 3$ and 5
will be sufficient, clearly in the accessible domain of the AMBER beam
combiner (Petrov et al.~\cite{petrov00}) with the 1.8\,m Auxiliary Telescopes
(Koehler et al.~\cite{koehler02}). The creation of a secondary network of calibrators should
therefore not be necessary.

It should be stressed that the present conclusions regarding magnitudes
are not limited to dwarf stars, as giants and supergiants
follow comparable surface brightness relations.
This means that the magnitude ranges defined above will be almost the same for other
luminosity classes. A decisive advantage of dwarfs is that for
the same apparent magnitude, they will be much closer than the more luminous classes,
and therefore significantly less affected by interstellar extinction.

\subsection{Example of diameter prediction \label{51peg_diam}}
 
As a practical application, we have chosen the two stars 51\,Peg\,A (\object{HD\,217014})
and HD\,209458\,A. The former hosts the first planet discovered around a solar type star
(Mayor \& Queloz~\cite{mayor95}), and the latter presents planetary transits 
(Charbonneau et al.~\cite{charbonneau00}). We selected these two stars because they
have been observed extensively using different techniques and
did not show any large amplitude photometric variability. They therefore represent good
examples of stable, well known stars, and are ideal candidates for the prediction of
their angular size using the SB relations determined in the present paper.

Table~\ref{table_predict} presents the predicted angular diameters of the two stars
for the $(V,\,V-K)$ version of the SB relation.
The $m_V$ magnitudes are from {\it Hipparcos} (Perryman et al.~\cite{hip}) for both stars,
with an arbitrary error bar of $\pm 0.01$, while
the $K$ band infrared magnitudes were taken from
Ducati et al.~(\cite{ducati02}) for 51\,Peg\,A, and from
the 2MASS catalogue (Cutri et al.~\cite{cutri03}) for HD\,209458\,A.
\begin{table}
\caption{Photometry (upper part) and predicted
limb darkened angular diameters $\theta_{\rm LD}$
(lower part) of the planet-hosting stars 51\,Peg\,A and HD\,209458\,A.}
\label{table_predict}
\begin{tabular}{lcc}
\hline
 & 51\,Peg\,A & HD\,209458\,A \\
\hline \noalign{\smallskip}
$m_V$ & $5.50 \pm 0.01$ & $7.65 \pm 0.01$ \\
$m_K$ & $3.97 \pm 0.01$ & $6.31 \pm 0.03$ \\
\hline
$\theta_{\rm LD} (K,\ V-K)$ & $0.689 \pm 0.011$\,mas & $0.228 \pm 0.004$\,mas \\
\hline
\end{tabular}
\end{table}

For 51\,Peg\,A, we obtain a predicted angular diameter of
$\theta_{\rm LD} = 0.689 \pm 0.011$~mas.
The corresponding value for HD\,209458\,A is
$\theta_{\rm LD} = 0.228 \pm 0.004$~mas.
These angular sizes can be translated into linear radii
using the {\it Hipparcos} parallaxes (Perryman et al.~\cite{hip}),
$\pi_{\rm 51\,Peg\,A} = 65.10 \pm 0.76$~mas and
$\pi_{\rm HD\,209458\,A} = 21.24 \pm 1.00$~mas.
We obtain
$R_{\rm 51\,Peg\,A} = 1.138 \pm 0.023 \,R_{\odot}$ and
$R_{\rm HD\,209458\,A} = 1.154 \pm 0.059\,R_{\odot}$.
HD\,209458\,A is a particularly interesting object, as Brown et al.~(\cite{brown01})
have been able to estimate directly its linear radius through the
deconvolution of the light curve of the transit. They obtain a value of
$R_{\rm HD\,209458\,A} = 1.146 \pm 0.050\,R_{\odot}$, in remarkable agreement
with our $(K,\ V-K)$ prediction. The bulk of the $\pm 5\%$ uncertainty
comes from the error in the {\it Hipparcos} parallax, the relative
error in the angular size being only $\pm 2\%$.

The direct measurement of the angular diameter of 51\,Peg\,A is within the
capabilities of the existing very long baseline interferometers (several hundred
meters), but this is not true for HD\,209458\,A. Its 0.228\,mas size would require
baselines of more than 800\,m to be resolved at visible wavelengths (several kilometers
in the infrared).
Such baselines are not presently available or scheduled.
And even so, the calibration of these observations would be
extremely difficult, as the calibrator would have to be very faint. More generally,
carefully calibrated surface brightness relations are currently the
only method to estimate precisely ($\pm 1\%$) the angular size of solar type stars
fainter than $m_V = 7$.

\section{Conclusion}

The laws that we established between the angular size and
broadband colors (or effective temperature) are strictly empirical.
Our best relations present a very small intrinsic dispersion, down to less than 1\,\%.
They can be used to predict the angular sizes of A0--M2 dwarfs and A0--K0
subgiants from simple, readily available broadband photometry.
On the one hand, Gray et al.~(\cite{gray03}) have recently published an extensive survey of the
spectral properties of nearby stars within 40\,pc, including estimates of their effective temperatures.
On the other hand, several large catalogues (2MASS, DENIS,...) provide high precision
magnitudes of these stars in the infrared. From the cross-comparison of these
sources, the SB relations determined in the present paper make it possible to
assemble a catalogue of calibrators for interferometry that will be practically unaffected
by interstellar extinction, multiplicity or circumstellar material biases.
These resulting angular diameter predictions will provide a reliable basis
for the calibration of long-baseline interferometric observations.

\begin{acknowledgements}
P.K. acknowledges partial support from the European Southern
Observatory through a post-doctoral fellowship.
D.S. acknowledges the support of the Swiss FNRS.
This research has made extensive use of the SIMBAD and VIZIER
online databases at CDS, Strasbourg (France).
\end{acknowledgements}

\end{document}